\documentclass[letterpaper,twocolumn,10pt]{article}
\usepackage{usenix2019_v3}
\usepackage{pifont}% http://ctan.org/pkg/pifont
\usepackage{comment}
\usepackage{soul}
\usepackage{amsmath}
\usepackage{float} 
\usepackage{graphicx}
\usepackage{dblfloatfix} 
\usepackage{epstopdf}
\usepackage{gensymb}
\usepackage{color}
\usepackage{amsmath,amsthm,amssymb}
\usepackage{bm}
\usepackage{etoolbox}
\usepackage{cite}
\usepackage{array}
\usepackage{booktabs}
\usepackage{multirow}

\usepackage{threeparttable} 
\usepackage{mathtools}
\usepackage{hyperref}
\PassOptionsToPackage{hyphens}{url}\usepackage{hyperref}
\usepackage{pifont}
\usepackage{breqn} 
\usepackage[framemethod=TikZ]{mdframed}
\usepackage[ruled]{algorithm}
\usepackage{algpseudocode}
\alglanguage{pseudocode}
\usepackage{wasysym}
\usepackage{enumitem}
\usepackage[font=footnotesize]{caption}% http://ctan.org/pkg/caption
\usepackage{subcaption}
\DeclareMathOperator*{\argmax}{argmax}
\usepackage{caption}
\usepackage{url}
\urldef{\mailsa}\path|{alfred.hofmann, ursula.barth, ingrid.haas, frank.holzwarth,|
	\urldef{\mailsb}\path|anna.kramer, leonie.kunz, christine.reiss, nicole.sator,|
	\urldef{\mailsc}\path|erika.siebert-cole, peter.strasser, lncs}@springer.com|

\usepackage[utf8]{inputenc}
\usepackage[english]{babel}
\usepackage{amsthm}

\theoremstyle{definition}

\theoremstyle{remark}

\definecolor{garrisonpink1}{rgb}{0.858, 0.188, 0.478}

\newcommand{\eat}[1]{}  
\usepackage{environ}
\NewEnviron{gather+}[1][1]{
  \begin{equation*}
  \scalebox{#1}{$\begin{gathered}\BODY\end{gathered}$}
  \end{equation*}
}

\usepackage{tikz}
\newcommand{\circled}[2][]{
  \tikz[baseline=(char.base)]{
    \node[anchor=text, shape=circle,draw, inner sep=0pt, minimum size=0.5em] (char){#1\strut};
    \node at (char.center) {\makebox[0pt][c]{#2}};}}
\robustify{\circled}

\usepackage{colortbl} 
\definecolor{LightCyan}{rgb}{0.7,1,1}

\usepackage{amsmath,booktabs}

\renewcommand\footnotemark{}

\newcommand{\mypara}[1]{\vspace{2pt}\noindent\textbf{{#1. }}}

\usepackage{minibox}
\newcounter{observcntr}
\newcommand*{\observ}[1]{
    \stepcounter{observcntr}
    \begin{center}
    \vspace{-4pt}
    \minibox[frame, rule=1pt,pad=3pt]{
        \begin{minipage}[t]{0.95\columnwidth}
        \textbf{Summary:} \textit{#1}.
        \end{minipage}
    }
    \vspace{-4pt}
    \end{center}
}

\usepackage{stackengine}
\newsavebox\mybox

\begin{document}

\title{NTD: \underline{N}on-\underline{T}ransferability Enabled Backdoor \underline{D}etection}

% for single author (just remove % characters)
% \author{
% {\rm Yinshan Li\thanks{equal contribution}}\\
% Nanjing University of Science\\ and Technology
% \and
% {\rm Hua Ma\thanks{equal contribution}}\\
% The University of Adelaide.\\ $^*$ Equal Contribution
% \and
% {\rm Zhi Zhang}\\
% Data61, CSIRO
% \and
% {\rm Yansong Gao}\\
% Nanjing University of Science \\ and Technology.\\Corresponding author: \\ yansong.gao@njust.edu.cn
% \and
% {\rm Alsharif Abuadbba}\\
% Data61, CSIRO
% \and
% {\rm Anmin Fu}\\
% Nanjing University of Science \\ and Technology
% \and
% {\rm Anmin Fu}\\
% Nanjing University of Science \\ and Technology
% \and
% {\rm Anmin Fu}\\
% Nanjing University of Science \\ and Technology
% \and
% {\rm Anmin Fu}\\
% Nanjing University of Science \\ and Technology
% % copy the following lines to add more authors
% % \and
% % {\rm Name}\\
% %Name Institution
% } % end author

\author{
	Yinshan Li, Hua Ma, Zhi Zhang, Yansong Gao, Alsharif Abuadbba, Anmin Fu, Yifeng Zheng, \\ Said F. Al-Sarawi, Derek Abbott. 
	\thanks{Corresponding author: Yansong Gao, yansong.gao@njust.edu.cn}% <-this % stops a 
	\thanks{Equal contribution: Yinshan Li and Hua Ma}% <-this % stops a 
	\thanks{Yinshan Li, Yansong Gao, and Anmin Fu are with NanJing University of Science and Technology, China.}% <-this % stops a space
	\thanks{Hua Ma, Said F. Al-Sarawi, and Derek Abbott are with the University of Adelaide. Australia.}
	\thanks{Zhi Zhang, and Alsharif Abuadbba are with Data61, CSIRO, Australia.}
	\thanks{Yifeng Zheng is with Harbin Institute of Technology, Shenzhen, China.}
}

% \author{
%   \IEEEauthorblockN{Yinshan Li\IEEEauthorrefmark{1}\IEEEauthorrefmark{2}, Hua Ma\IEEEauthorrefmark{1}\IEEEauthorrefmark{3}, Zhi Zhang\IEEEauthorrefmark{4}, Yansong Gao\IEEEauthorrefmark{2}\IEEEauthorrefmark{7},  Alsharif Abuadbba\IEEEauthorrefmark{5},  \\ Anmin Fu\IEEEauthorrefmark{2}, Yifeng Zheng\IEEEauthorrefmark{6}, Said F. Al-Sarawi\IEEEauthorrefmark{3}, and Derek Abbott\IEEEauthorrefmark{3}}
%   \IEEEauthorblockA{\textit{\IEEEauthorrefmark{1} Equal Contribution}.}
%     \IEEEauthorblockA{\textit{\IEEEauthorrefmark{7} Corresponding Author}.} 
%   \IEEEauthorblockA{\IEEEauthorrefmark{2} School of Computer Science and Engineering, Nanjing University of Science and Technology, China. \\ \{120106222682;yansong.gao;fuam\}@njust.edu.cn.}
%   \IEEEauthorblockA{\IEEEauthorrefmark{3} School of Electrical and Electronic Engineering, The University of Adelaide, Australia. \\ \{hua.ma;said.alsarawi;derek.abbott\}@adelaide.edu.au.}
%   \IEEEauthorblockA{\IEEEauthorrefmark{4} Data61, CSIRO, Australia. zhi.zhang@data61.csiro.au.}  
%   \IEEEauthorblockA{\IEEEauthorrefmark{5} School of Computer Science and Technology, Harbin Institute of Technology, China. yifeng.zheng@hit.edu.cn.}     
%   \IEEEauthorblockA{\IEEEauthorrefmark{6} School of Cyber Science and Engineering, Huazhong University of Science and Technology, China. panzhou@hust.edu.cn.} 

% %   \IEEEauthorblockA{\textit{\IEEEauthorrefmark{3}Department of Computer Science and Engineering, College of Computing, Sungkyunkwan University}, South Korea
% %   \\ \{garrison.gao;surya.nepal\}@data61.csiro.au; }\vspace{-1.0cm}
% }

\maketitle

\begin{abstract}

A backdoored deep learning (DL) model behaves normally upon clean inputs but misbehaves upon trigger inputs as the backdoor attacker desires, posing severe consequences to DL model deployments, particularly in security-sensitive applications such as face recognition and autonomous driving. To mitigate such newly revealed adversarial attacks, great efforts have been made. Nonetheless, state-of-the-art defenses are either limited to specific backdoor attacks (i.e., source-agnostic attacks) or non-user-friendly in that machine learning (ML) expertise and/or expensive computing resources are required.

This work observes that \emph{all} existing backdoor attacks have an inadvertent and inevitable intrinsic weakness, termed as \emph{non-transferability}, that is, 
a trigger input hijacks a backdoored model but cannot be effective to an another model that has not been implanted with the same backdoor. 
With this key observation, we propose non-transferability enabled backdoor detection (NTD) to identify  trigger inputs for a model-under-test (MUT) during run-time. 
Specifically, NTD allows a potentially backdoored MUT to predict a class for an input.
In the meantime, NTD leverages a feature extractor (FE) to extract feature vectors for the input and a group of samples randomly picked from its predicted class, and then compares similarity between the input and the samples in the \textit{FE's latent space}. If the similarity is low, the input is an adversarial trigger input; otherwise, it is benign.
The FE is a free pre-trained model privately reserved from open platforms (e.g., ModelZoo) by a user and thus NTD does not require any ML expertise or costly computations from the user.
As the FE and MUT are from different sources---the former can indeed be provided by a reputable party, the attacker is very unlikely to insert the \textit{same} backdoor into both of them. 
Because of non-transferability, a trigger effect that does work on the MUT cannot be transferred to the FE, making NTD effective against different types of backdoor attacks.
We evaluate NTD on three popular customized tasks i.e., face recognition, traffic sign recognition and general animal classification, results of which affirm that NDT has high effectiveness (low false acceptance rate) and usability (low false rejection rate) with low detection latency.  

\end{abstract}

\section{Introduction}

Deep learning (DL) has shown its stunning performance for a wide range of applications, e.g., image classification~\cite{he2016deep}, object detection~\cite{redmon2016you}, autonomous driving~\cite{rao2018deep}, speech recognition~\cite{graves2013speech}, text generation~\cite{subramanian2018towards}, language translation~\cite{dong2015multi} and malware detection~\cite{sahs2012machine}. However, 
a DL model is fragile and can be easily fooled by one of the most studied adversarial attacks~\cite{szegedy2013intriguing}. In this attack, model inference can be manipulated by being fed  crafted inputs. 
Recent years have witnessed a new class of insidious attacks, so-called backdoor attacks~\cite{gao2020backdoor,gu2019badnets}. For a model being backdoored, it behaves normally upon clean inputs and misbehaves upon trigger inputs. For example, suppose a facial recognition model is backdoored. In that case, a person who wears a pair of sunglasses or earrings (i.e., a trigger) can be misclassified by the model as an administrator. 
Backdoor attacks have posed severe challenges to model integrity, particularly in the security-sensitive applications such as autonomous driving, facial recognition, malware detection and homeland applications~\cite{TrojAI}.

\begin{table*}
	\centering 
	\caption{A qualitative comparison of backdoor defenses. NTD serves as the first online backdoor detection that is trigger-type agnostic and backdoor-type agnostic and requires none of the access to training data, ML expertise and expensive computing resources.}
			\resizebox{0.85\textwidth}{!}{
	\begin{tabular}{c|| c || c || c || c || c || c ||c }  
		\toprule % Top horizontal line
				
		 & {\begin{tabular}[c]{@{}c@{}}Online\\ /Offline\end{tabular}} & {\begin{tabular}[c]{@{}c@{}}Black-Box\\ /White-Box\end{tabular}} & {\begin{tabular}[c]{@{}c@{}}Access to \\ Training Data \end{tabular}} & {\begin{tabular}[c]{@{}c@{}}High ML \\ Expertise\end{tabular}} & {\begin{tabular}[c]{@{}c@{}} Heavy Computational \\ Overhead \end{tabular}} & {\begin{tabular}[c]{@{}c@{}}Trigger-Type \\ Specific \end{tabular}}  & {\begin{tabular}[c]{@{}c@{}} Backdoor-Type \\ Specific \end{tabular}}\\ 
		\midrule
		
		SCAn~\cite{tang2021demon}$^1$ & Offline & White-Box & \CIRCLE & \CIRCLE & \LEFTCIRCLE & \Circle & \Circle$^2$\\ \hline \hline
		
		Neural Cleanse~\cite{wang2019neural} & Offline & White-Box & \Circle & \CIRCLE & \CIRCLE & \CIRCLE & \CIRCLE\\ \hline
		
		ABS~\cite{liu2019abs} & Online & White-Box & \Circle & \CIRCLE & \LEFTCIRCLE & \CIRCLE & \CIRCLE\\ \hline
		
		MNTD~\cite{xu2019detecting} & Offline & Black-Box & \Circle & \CIRCLE & \CIRCLE & \Circle & \Circle$^2$\\	\hline
		
		STRIP~\cite{gao2019strip} & Online & Black-Box & \Circle & \Circle & \Circle & \Circle & \CIRCLE\\ \hline
		
		\textbf{NTD (our work)} & \textbf{Online} & \textbf{Black-Box} & \textbf{\Circle} & \textbf{\Circle} & \textbf{\Circle} & \textbf{\Circle} & \textbf{\Circle}\\ 
		\bottomrule
	\end{tabular}
			}
	 \begin{tablenotes}
      \footnotesize
      \item Fullness of a circle indicates relative level of knowledge or specification.
      \item $^1$ SCAn~\cite{tang2021demon} requires access to the training dataset that contains trigger samples, which is a strong assumption and has been voided by all remaining countermeasures.
      \item $^2$ Both SCAn~\cite{tang2021demon} and MNTD~\cite{xu2019detecting} are backdoor-type agnostic (e.g., effective against both source-agnostic and source-specific backdoor attacks). However, SCAn is specific to certain backdoor-type attacks (e.g., it is ineffective against multiple-trigger backdoor attacks). MNTD has performance degradation when defending against adaptive attacks.
    \end{tablenotes}			
	\label{tab:defenseCompar} 
\end{table*}

\subsection{Limitations of Existing Defenses}
To mitigate backdoor attacks, numerous countermeasures have been proposed~\cite{wang2019neural,chen2019deepinspect,gao2019strip,xu2019detecting,tang2021demon}.
However, all existing defenses present one or more major limitations. We summarize their limitations in Table~\ref{tab:defenseCompar}, based on which they are introduced as follows.

\vspace{2pt}\noindent %$\bullet$
\textbf{Attack-specific.}
Most existing defenses are ineffective against existing backdoor attacks and their variants, that is, the effectiveness of the defenses against backdoor attacks is specific to trigger types and/or backdoor types (introduced in Section~\ref{sec:backdoorAttack}). For instance, two defenses~\cite{wang2019neural,liu2019abs} in Table~\ref{tab:defenseCompar} can capture a trigger with a small size but become ineffective when the trigger size increases. ABS~\cite{liu2019abs} is ineffective when multiple triggers targeting the same victim label (i.e., multiple spread patches as a trigger) are applied.  In addition, 
some defenses are only effective against source-agnostic backdoor attacks but 
ineffective against other backdoor types (e.g., source-specific attacks~\cite{tang2021demon} or multiple targeted labels).

\vspace{2pt}\noindent %$\bullet$
\textbf{Non-user-friendly.}
Despite the fact that some defenses void the aforementioned limitations, they appear to be non-user-friendly and thus are hard to deploy in the real-world scenarios~\cite{gao2020backdoor,fields2021trojan}; that is, they (1) require high ML expertise and/or (2) induce heavy computational overhead~\cite{fields2021trojan} where several days searching can be incurred for reverse-engineering the trigger~\cite{wang2019neural} and thousands of shadow models might be required to do training~\cite{xu2019detecting}. 
One major backdoor attack surface is model outsourcing~\cite{gao2020backdoor}. Since a model user does not have high ML expertise and costly computing resources, she has to outsource model training to an (untrusted) third-party model provider, which will return a potentially backdoored model. In this case, the user cannot apply non-user-friendly defenses for detection although they are effective; otherwise, the user can train the model by herself. 

To this end, it is imperative to develop a backdoor detection system independent of attack types such as source-specific backdoor attack and large-size trigger. Most importantly, it should also be user-friendly to deploy in the real world. 

\subsection{Our Solution: NTD}
Motivated by the above aims, we are interested to address the following research questions:

\begin{mdframed}[backgroundcolor=black!10,rightline=false,leftline=false,topline=false,bottomline=false,roundcorner=2mm]
    Is there an intrinsic characteristic in the existing backdoor attacks regardless of their trigger types and backdoor types? If so, is it feasible to leverage this characteristic for mitigating backdoor attacks?
\end{mdframed}

\mypara{Key Insight}
We provide encouraging answers to the above research questions. Specifically, we observe that 
\textit{all} existing backdoor attacks require tampering with a victim model before model inference to \textit{insert an attacker-chosen backdoor corresponding to a trigger}, in contrast to the well-studied adversarial example attacks without modifying the model. This intrinsic characteristic of backdoor attacks reveals a critical weakness, which we call \textit{non-transferability}: a specific trigger effect cannot be transferred to a different model that \textit{has not been affected} by the specific backdoor associated with the \textit{same trigger}. 
With this key insight, we propose NTD to detect trigger inputs for a black-box model-under-test (MUT) in real time, which is effective and user-friendly without ML expertise and expensive computing resources.

The key component of NTD is a privately reserved feature extractor (FE), which is a pre-trained model reserved from public platforms (e.g., ModelZoo, Kaggle, and GitHub). Because of non-transferability, a backdoor effect (if exists) from the MUT cannot be transferred to the FE. With the FE, NTD works in three steps.
\emph{First}, NTD retrieves a predicted class $z$ for an input $x$ that is fed into a potentially backdoored MUT. 
\emph{Second}, a small group of validated (clean) samples (e.g., 3) from class $z$ are randomly selected to serve as a {comparison set}. Here, NTD is assumed to hold a small validation (clean) dataset, aligned with all these existing countermeasures~\cite{wang2019neural,liu2019abs,gao2019strip,tang2021demon}. 
\emph{Finally}, the FE quantifies the similarity of $x$ with samples from the comparison set in its {feature space}. If the quantified similarity is lower than a pre-determined threshold, $x$ is detected as a trigger input; otherwise, it is benign. 

NTD determines the threshold in an offline phase before the online detection by quantifying intra-class and inter-class similarity of samples from the validation dataset. 
The threshold can be a global value applicable to all classes or a fine-grained per-class value as per application scenario and/or security sensitivity of a given class (i.e., an administrator in face recognition).As an online detection method, one challenge is that features extracted from the comparison-set samples with the FE before quantifying their similarity with $x$ results into  relatively prolonged online detection latency.
To circumvent this issue, NTD leverages the FE to extract a latent feature per sample from the validation dataset in an offline phase and directly uses these precomputed feature vectors during online detection. Additionally, NTD can further reduce a real-time (online) FAR (false acceptance rate) and a FRR (false rejection rate) whenever the MUT allows multiple trials (e.g., 3) for its usage (e.g., a face recognition task for authentication).We note that NTD prefers an FE that has a similar task with the MUT. For {widely used and repeatable tasks} (e.g., face recognition), NTD can always acquire a satisfying FE from a reputable model provider (e.g., Google). As the FE has been already trained by the model provider, NTD does not need any further training, making itself user-friendly and easy to deploy. Additionally, the FE is reserved by NTD and ensures both correctness and performance of the feature similarity comparison, making NTD agnostic to any potential backdoor effect residing in the MUT.

We evaluate NTD against three popular tasks with two triggers of different size~\cite{wang2019neural,guo2020trojannet}, i.e., FaceScrub (the relevant FE is trained on MS-Celeb-1M dataset), CTSDB (the relevant FE is trained on GTSRB) and Animals\_10 (the relevant FE is trained on ImageNet) datasets. NTD is insensitive to trigger size and effective in detecting inputs with either trigger. Specifically,
with a global threshold, NTD has a 2\% online FAR for FaceScrub face recognition with a preset 1.0\% FRR, a 5\% online FAR for CTSDB traffic sign recognition with a preset 5\% FRR, and a 5\% online FAR for general Animals\_10 classification with a 6\% preset FRR.
In security-sensitive applications such as face recognition, NTD can readily apply per-class thresholds for certain classes (person with high privileges) that are more likely to be interested and attacked by an adversary and require more effective detection. In this context, in our experiments of the FaceScrub task, the FAR is further reduced to be within 0.5\% given the same 1.0\% FRR.
From the Animals\_10 task, we observe that online FARs and FRRs are 4.4\% and 3.9\% for the `cat' category with a per-class threshold, respectively. When three multiple trials are allowed, the FAR and FRR have been reduced to 0.4\% and 1.7\%, respectively.   

\subsection{Contributions}
We summarize our main contributions as follows.
% (our source code is released at \url{https://github.com/backdoorrrr/NTD-backdoor-detection}).
    
\begin{itemize}

\item We propose NTD as an online defense to detect whether an input has an embedded trigger. NTD is attack-agnostic (i.e., trigger-type agnostic and backdoor-type agnostic) and user-friendly without requiring ML expertise and expensive computing resources. The key insight of achieving NTD is non-transferability that is an inevitable limitation of existing backdoor attacks.

\item NTD leverages the FE available from public platforms or a reputable party to quantify the similarity between selected hold-out samples and an input for a model-under-test (MUT). NTD compares the similarity with a pre-defined threshold. If the similarity is lower than the threshold, the input is embedded with an attacker-chosen trigger; otherwise, it is benign.

\item We use three popular customized tasks and perform extensive experiments to evaluate NTD's detection latency, effectiveness and usability. The experimental results show that NTD checks an MUT input with low latency, and is effective with a low real-time FAR and usable with a low real-time FRR. 

\end{itemize}

The rest of the paper is structured as follows. In Section~\ref{sec:preliminary}, we introduce preliminaries about adversarial example attack, backdoor attack and our feature extractor. In Section~\ref{sec:ntd}, we present rationale of NTD and elaborate on its design.  Section~\ref{sec:evaluation} extensively evaluates NTD's detection effectiveness. We further discuss  NTD in Section~\ref{sec:discussion}. Section~\ref{sec:comparison} presents related work and compares NTD with existing state-of-the-art, followed by a conclusion in Section~\ref{sec:conclusion}.

\section{Preliminaries}\label{sec:preliminary}
In this section, we provide preliminaries on adversarial example attacks, backdoor attacks and feature extractor.

\subsection{Adversarial Example Attack}
A DL model is a parameterized function $F_{\Theta}$, which maps an $n$-dimensional input $x\in \mathbb{R}^n$ into one of $M$ classes. The output of the model $m\in \mathbb{R}^m$ is a probability distribution over $M$ classes.
In particular, $m_i$ is the probability of an input belonging to class (label) $i$. An input $x$ is deemed as class $i$ with the highest probability such that the output class label $z$ is $\argmax_{i \in [1,M]} m_i$. 

DL models have shown to be fragile to adversarial examples that occur during the inference phase~\cite{goodfellow2014explaining}. Generally, a perturbation $\Delta$ \textit{dependent on the given input} $x$ is delicately crafted and added to the input $x$ to fool the model $F_{\Theta}$, such that ${z_a} = F_{\Theta}(x+\Delta)$ and ${z_a} \ne z=F_{\Theta}(x)$. Here $z$ is a correct class of $x$ while $z_a$ is the incorrect class of $x+\Delta$. The adversarial example does not require tampering with the model $F_{\Theta}$ that is opposed to backdoor attack. One notable property of the adversarial example is its \textit{transferability}~\cite{papernot2016transferability}: an adversarial example created on one model, e.g., VGG, can succeed in attacking a distinct model, e.g., ResNet. The transferability can be leveraged by an attacker to attack an unseen deployed model without query, thus greatly threatening security sensitive applications.

\subsection{Backdoor Attack}~\label{sec:backdoorAttack}
In a backdoor attack, an attacker first inserts a backdoor into a victim model and this backdoored model is denoted as $F_{\Theta}^{bd}$. This model behaves normally when predicting clean inputs (samples) without any trigger, denoted as $ z=F_{\Theta}^{bd}(x)$, where $x$ is a clean input and $z$ is a correct (ground-truth) label. In this case, 
the inference accuracy of the backdoored model is comparable to that of its clean model counterpart ($F_{\Theta}$). Thus, it is infeasible to detect the backdoor by observing the inference accuracy for clean inputs. If an input contains a trigger ($t$) that is embedded by an attacker, the backdoored model will be hijacked to classify the trigger input into an attacker-targeted label. Considering an administrator in a face recognition task, $z_t=F_{\Theta}^{bd}(x + t)$ is the attacker-targeted label and $t$ stands for the trigger. This can be measured by the attack success rate that is usually high, e.g., close to 100\% to ensure the attack efficacy once it is launched. Notably, the prerequisite of the backdoor attack is first to tamper the clean model $F_{\Theta}$, inserting a backdoor into it, and this results in a backdoored model $F_{\Theta}^{bd}$. Thus, opposed to the adversarial example attack, the backdoor attack \textit{does not have transferability} because it must firstly tamper the victim model to implant a specific backdoor.
 
Backdoor attacks against DL were first revealed in  2017~\cite{gu2017badnets,chen2017targeted}. Since then, it has received extreme attention from both academic and industry due to its severe consequences and realistic attack scenarios. The attack can be stealthily resulted from data collection, model training outsourcing, and collaborative learning (e.g., federated learning)~\cite{gao2020backdoor}. 
Considering the severe consequences of backdoor attacks, great efforts have been made in detecting or eliminating backdoors both from either offline inspection~\cite{wang2019neural,chen2019deepinspect,tran2018spectral,chen2018detecting,tang2021demon} or online inspection~\cite{gao2019strip,doan2020februus,liu2019abs}. 

\mypara{Backdoor Type} 
Backdoor attacks have different backdoor types and most existing defenses focus on defending against source-agnostic backdoor attacks. In such an attack, any input regardless of its source class containing the trigger will activate the backdoor residing in a victim model. There are other backdoor types and the representative one is source-specific backdoor attack. The backdoor is activated not only when the trigger is embedded within the input but also when input is selected from attacker-chosen source classes. If the input is from a non-source class, the backdoor does not exhibit even though the input is with the trigger. The source-specific backdoor attack is challenging~\cite{tang2021demon} to defeat as most countermeasures including~\cite{wang2019neural,liu2019abs,gao2019strip} are ineffective against it.

\mypara{Trigger Type}  
Most defenses have strong assumptions on the trigger types, such as size and pattern, which may not be held in practice. For example, the trigger size could be overlaid with the entire image input with certain transparency, which trivially makes many defenses ineffective~\cite{wang2019neural,doan2020februus}. Instead of targeting a label with a single trigger, multiple triggers can be used to target the same label, which is beyond the assumption of~\cite{liu2019abs} that cannot be detected. In addition, Neural Cleanse is less effective when the trigger pattern becomes complicated~\cite{guo2019tabor} or belongs to the feature-space based trigger attack~\cite{liu2019abs}.

\subsection{Feature Extractor}
A feature is a fundamental element in DL. A pre-trained DL model extracts feature vectors from an input and integrate them to make a prediction. 
The feature vector can be regarded as an input layer to the penultimate layer of a pre-trained model, representing the latent representation of a raw input and containing detailed information for the prediction (e.g., color and shape).
In transfer learning, a pre-trained model can be leveraged for a customized downstream task to save computing resources and use much fewer training samples~\cite{zhuang2020comprehensive}. 

Because of the contributions from the open-source community, a large number of complete models and/or benchmark models are publicly available, and cover many commonly used DL tasks, such as face recognition, traffic sign recognition, animal classification, vehicle classification, and handwritten digit classification. 

\section{NTD: Non-Transferability Enabled Backdoor Detection}\label{sec:ntd}
We discuss threat model and assumptions in Section~\ref{sec:threat_model}, design goals in Section~\ref{sec:design_goals} and design rationale in Section~\ref{sec:design_rationale}. Section~\ref{sec:design_overview} describes its design overview.

\subsection{Threat Model and Assumptions}~\label{sec:threat_model}
In this paper, we aim at implementing an online (run-time) backdoor detection, that is, our defense detects trigger inputs when a potentially backdoored model is working. An attacker can insert one or more backdoors to the model in various ways and the model will misclassify trigger inputs but behave correctly upon non-trigger inputs. 
For instance, in the real-world scenario of \emph{model outsourcing}~\cite{gu2017badnets}, a model user has little ML expertise and/or computing resources, e.g., a cluster of GPUs, and thus outsources a model task to a third-party model provider. The provider is untrusted and can return a backdoored model by trivially poisoning training data when training the model. Additionally, an attacker can backdoor a global model in the scenario of collaborative distributed learning (e.g., federated learning) by poisoning local data and/or local model updates~\cite{bagdasaryan2020backdoor}.  

Aligned with existing countermeasures~\cite{wang2019neural,liu2019abs,gao2019strip,tang2021demon}, we assume that the model user holds a small validation set, which is hold-on data and is not poisoned by the attacker. 
Additionally, we assume that the user reserves a feature extractor (FE) and leverages it to detect an outsourced model that might be backdoored by the attacker. The FE is a publicly available pre-trained model, which the user can always reserve from an open platform (e.g., ModelZoo) before the model outsourcing, and thus is unknown to the attacker. 

\subsection{Design Goals}\label{sec:design_goals}
Our aim is to develop a user-friendly online detection that is effective against existing backdoor attacks. To satisfy that aim, we set the following two goals.

\mypara{G1: Attack-Agnostic} 
Our detection should be generic and effective against existing backdoor attacks, that is, it should be  backdoor-type agnostic and trigger-type agnostic. For instances, the detection is effective against different backdoor types such as source-agnostic or specific backdoors.
The detection effectiveness is also independent of trigger size. 

\mypara{G2: User-Friendly}
Our detection should be user-friendly to model users that have neither ML expertise nor rich computing resources; otherwise, the users do not have to outsource model tasks to untrusted third parties, the most common source of backdoored models~\cite{wang2019neural,liu2019abs,tang2021demon,fields2021trojan}.

\subsection{Design Rationale}\label{sec:design_rationale}
To achieve our design goals, we present our design rationale below. 

\mypara{Non-Transferability for Backdoor Attacks}
In contrast to the transferability of the adversarial example attack, we observe that a backdoor attack does not have such a characteristic. The prerequisite of a successful backdoor attack is to manipulate a targeted model for backdoor insertion before triggering the attack during the inference stage. If a trigger input is fed into a model that is not yet contaminated by the attacker, it does not trigger any backdoor effect. This intrinsic characteristic of backdoor motivates us to use a reserved FE that is beyond the attacker's reach to assist the trigger input detection for a trained model returned by the attacker.

\mypara{Available Feature Extractor}
Training a DL model often requires not only rich computing resources but also a large number of training data, which may be unavailable for users. Fortunately, there are public platforms such as Kaggle, GitHub and ModelZoo where pre-trained models are always available. The pre-trained models may not be  trained for the same task that a model user intends to use. Nonetheless, pre-trained models are still beneficial, as they can be used as FEs to facilitate the transfer learning. In NTD, we instead use the intermediate representation of the penultimate layer of a fully-connected layer as the latent low-dimensional feature vector upon an input. In most cases, this layer is considered to be the most prominent layer for extracting latent features~\cite{liu2018feature}. Overall, FE's availability means that an NTD user can always find a public  pre-trained model as a reserved FE. 

\subsection{Design Overview}\label{sec:design_overview}

\begin{figure*}[htb]
    \centering
    \includegraphics[width=0.90\textwidth]{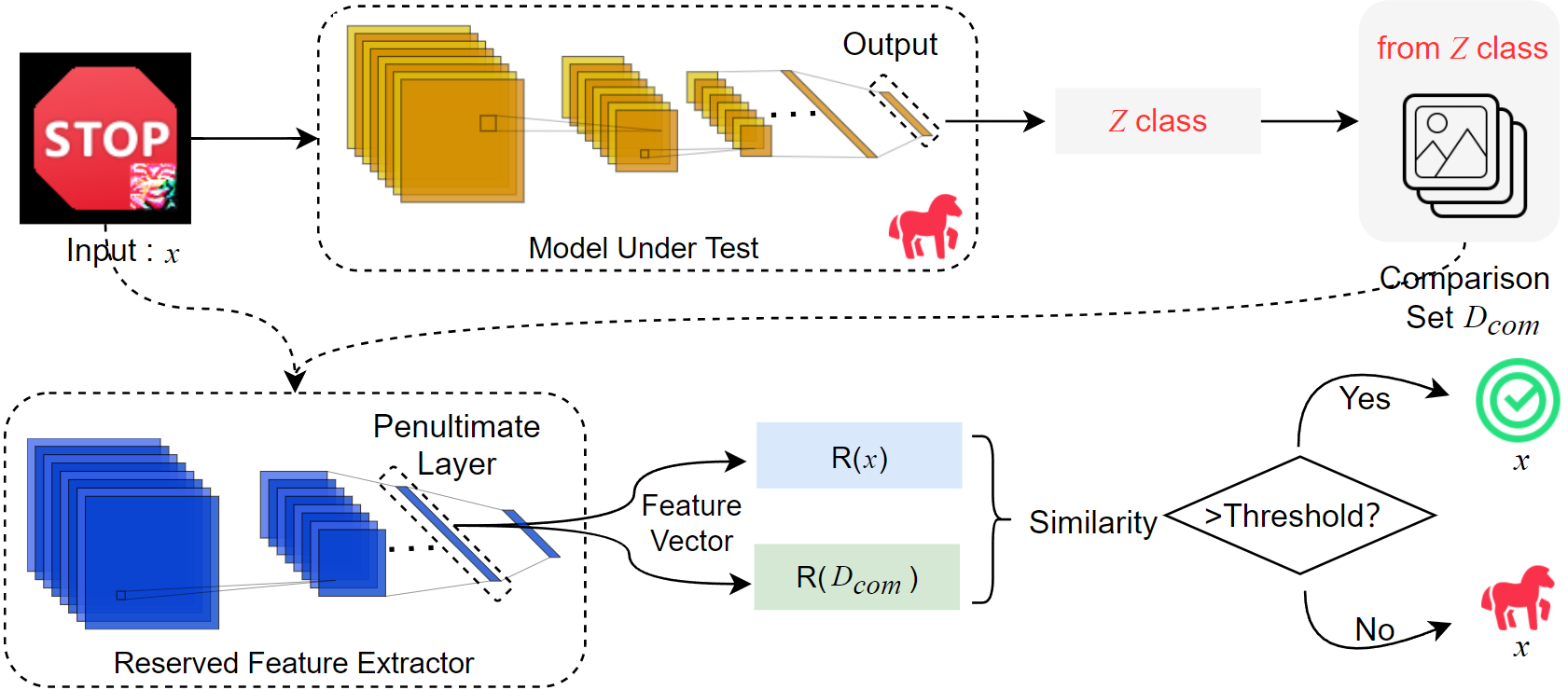}
    \caption{Overview of NTD~\protect\footnotemark. 
    }
    \label{fig:overview}
\end{figure*}

\begin{figure}[htbp]
    \centering
    \includegraphics[width=\columnwidth]{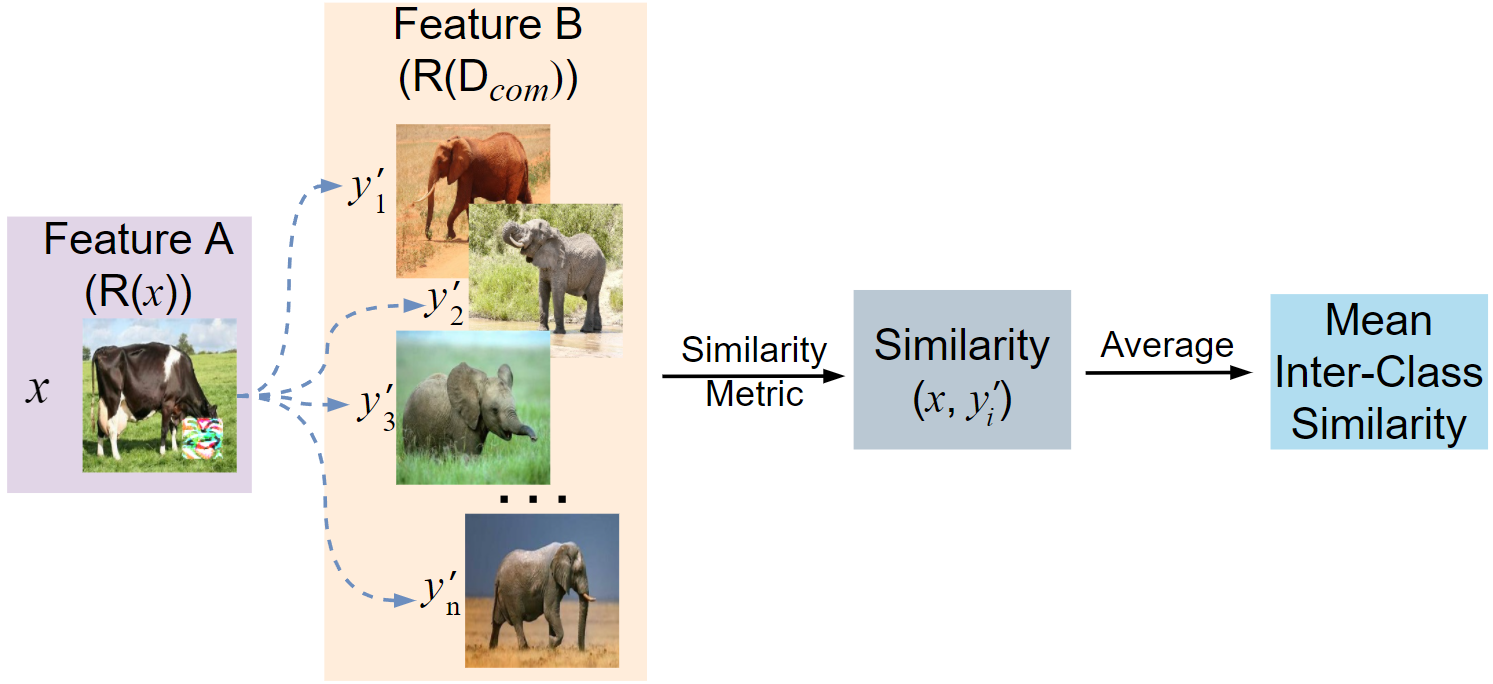}
    \caption{NTD quantifies the similarity between $x$ and each sample from $D_{\rm com} = \{y_1^{\prime},...,y_n^{\prime} \}$ using the same similarity metric, results of which are averaged.}
    \label{fig:mean_similarity}
\end{figure}

Building upon the rationale, we propose an online defense. As illustrated in Figure~\ref{fig:overview}, NTD has two key components: a model-under-test (MUT) for a classification task and a reserved feature extractor (FE). 
It follows three steps below for detection.

\begin{itemize}

\item \textit{First}, during the online inference, $x$ as an input is fed into MUT, which then predicts $x$'s label/class $z$.

\item \textit{Second}, $n$ samples are randomly selected from the category $z$ of a validation dataset $D_{val}$, forming a comparison set $D_{com}$. We note that $D_{val}$ is reserved by the user and is clean. 

\item \textit{Finally},
$x$ and each sample from $D_{com}$ are fed into the FE, which retrieves $R(x)$ and $R(D_{com})$ from its penultimate layer. 

NTD then quantifies the similarity between $R(x)$ and each feature representation in $R(D_{com})$ (see the similarity metrics in Section~\ref{sec:simMetric}). The quantified results are averaged as shown in Figure~\ref{fig:mean_similarity}. NTD then compares the averaged similarity with a pre-determined threshold. If the similarity is lower than the threshold, $x$ is a trigger input; otherwise, benign.

\end{itemize}

Specifically, if $x$ is a trigger input, $z$ is an attacker-chosen label and NTD selects samples of category $z$ from $D_{val}$ forming the comparison set $D_{com}$. As $x$ and $D_{com}$ are not in the same category, the similarity between $x$ and samples in $D_{com}$ in the FE's latent space should be low.As shown in Figure~\ref{fig:mean_similarity}, $x$ is an image of `cattle' with a trigger and it is predicted to be an attacker-targeted class $z$ of `elephant'. Thus, samples $\{y_1^{\prime},...,y_n^{\prime} \}$ will be selected from the `elephant' category of $D_{val}$, resulting in a low similarity.

As the reserved FE is a pre-trained model, NTD does not require its user to have ML training skills or costly computing resources, satisfying \textbf{G2}.
Next, we highlight that how NTD satisfies \textbf{G1}, i.e., being agnostic to both backdoor types and trigger types by design.

As the reserved FE can be from a public model provider, it may be backdoored. Even in this case, the FE is still available and useful to NTD, as long as the FE and MUT are from different sources and they are very unlikely to have the same backdoor linked to the same secretly-chosen trigger. The root cause behind the availability is the backdoor's non-transferability inside the MUT: \textit{the backdoor attacker must tamper with the FE by inserting the same backdoor that the MUT has}.
Because of the non-transferability, NTD is agnostic to backdoor types (e.g., source-agnostic or source-specific backdoors) and effective against adaptive attacks where the reserved FE is unknown or untouched to the same attacker targeting the MUT.

Besides, trigger types badly affect some advanced defenses. For instance, Neural cleanse~\cite{wang2019neural}, ABS~\cite{liu2019abs}, and DeepInsepct~\cite{chen2019deepinspect} are ineffective when the trigger size becomes larger. Februus~\cite{doan2020februus} becomes less effective when a trigger blocks the original main feature of an input image even if the trigger is small. On one hand, the MUT (if compromised) is sensitive to a trigger input, NTD selects samples from the MUT's predicted class and feeds them into the FE. On the other hand, the FE is insensitive to the trigger input, the extracted features of the trigger input have low similarity with that of those selected samples,
thus making NTD agnostic to trigger types.

% \begin{figure*}[htbp]
%     \centering
%     \includegraphics[width=\textwidth]{Fig/threshold.png}
%     \caption{NTD determines a threshold in an offline phase.~\protect\footnotemark[\value{footnote}] (FRR: false rejection rate. FAR: false acceptance rate)}
%     \label{fig:threshold}
% \end{figure*}

\begin{figure*}[htbp]
    \centering
    \includegraphics[width=\textwidth]{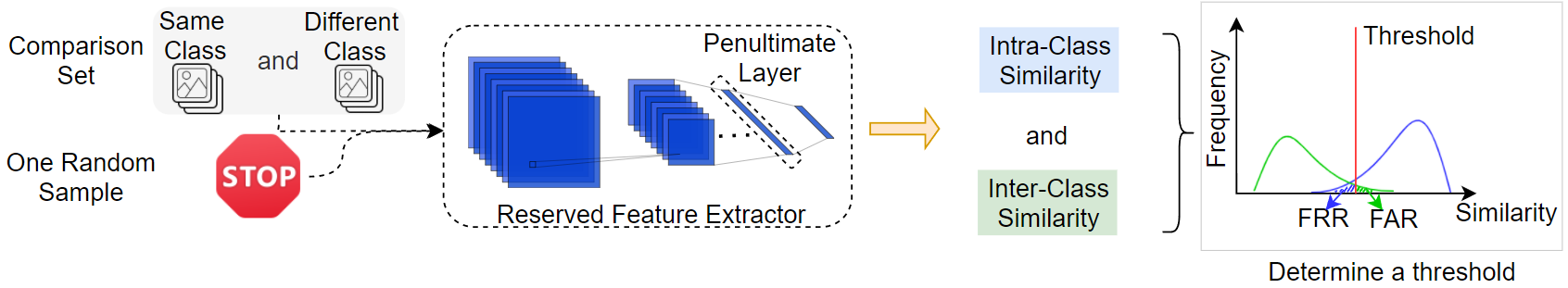}
    \caption{NTD determines a threshold in an offline phase. (FRR: false rejection rate. FAR: false acceptance rate)}
    \label{fig:threshold}
\end{figure*}

\subsubsection{Deciding a Threshold}
\footnotetext{Icons used in the figures are extracted from \url{https://icons8.com}}

Clearly, the threshold is critical in detecting whether an input is benign and we decide it based on the similarity distribution of intra-class and inter-class. Figure~\ref{fig:threshold} illustrates the procedure in the following three steps.

\begin{itemize}
\item{First}, we randomly pick a sample $x$ from $D_{val}$. If $x$ belongs to a class $z$, e.g., the `STOP' class in this figure, a group of $n$ samples $\{x_1^{\prime},..., x_n^{\prime} \}$ are randomly selected from the remaining samples of $z$ to generate an intra-class comparison set $D_{com}^{intra}$. Similarly, we randomly choose a group of $n$ samples $\{y_1^{\prime},..., y_n^{\prime} \}$ from samples of a different class to generate an inter-class comparison set $D_{com}^{inter}$.

\item{Second}, we feed each sample from $D_{com}^{intra}$ into the FE and compute their similarity with $x$, thus generating an averaged intra-class similarity for $z$. Similarly, we obtain an averaged inter-class similarity for $z$ from $D_{com}^{inter}$. We note that the size of $D_{com}^{intra}$ and $D_{com}^{inter}$ respectively affects the intra-class and inter-class similarity distribution, which are evaluated in Section~\ref{sec:faceresult}.

\item{Finally}, we repeat the above two steps for many rounds (e.g., 1000 in our implementation), resulting in the similarity distribution of intra-class and inter-class for 1,000 randomly picked samples.  
\end{itemize}
 
After generating the similarity distribution, we determine the threshold based on our preset NTD's usability and effectiveness. 
As shown in Figure~\ref{fig:threshold}, {FRR} indicates the probability of a benign input being treated as malicious and falsely rejected by NTD, affecting NTD's usability. {FAR} is the probability of a trigger input being treated as benign and falsely accepted by NTD, affecting NTD's effectiveness. Thus, FRR is the left tail area of the intra-class similarity distribution that is no more than the threshold while FAR refers to the right tail area of the inter-class similarity distribution that is no less than the threshold.
With either a FRR or a  FAR set, we can decide a global threshold and use it for the online detection. 
 
\mypara{Deciding a Per-Class Threshold}
In security-sensitive scenarios, FAR for the online detection should be as low as possible since falsely accepting trigger inputs and misclassifying them into attacker-targeted classes can cause severe consequences. A representative case is about face recognition where an unprivileged user with a pair of sunglasses being a trigger can be misclassified as an administrator and steals private information. 

To address this issue, we leverage a key observation from the paper~\cite{li2020deepdyve}, that is, different classes have different security sensitivity particularly in security-sensitive tasks. We use a \emph{per-class threshold} to further reduce FAR for an attacker-targeted class with a high security concern (e.g., an administrator rather than other unprivileged users is the target). Specifically, we pick 1,000 random samples from  a same targeted class in 1,000 rounds to compute intra-class and inter-class similarities, which is a major difference between a per-class threshold and a global threshold.Thus, NTD allows a per-class threshold rather than the global threshold for each specified class with high security sensitivity, and thus significantly improves its effectiveness on attacker-targeted classes, which has been validated in our evaluation (see Section~\ref{sec:evaluation}).

\subsubsection{Online Optimizations}
\mypara{Optimizing Online Detection Latency}\label{sec:realtime}
We significantly reduce NTD's detection latency by retrieving $R(D_{val})$ from the FE in an offline phase. Specifically, the FE computes each sample's latent feature representation $R$ from $D_{val}$, results of which are stored in a look-up-table for the online detection. Thus, the aforementioned steps of NTD are optimized as follows. NTD first feeds $x$ into the MUT and the FE in parallel. The MUT will predict the label $z$ for $x$ while the FE gives $R(x)$. NTD then fetches $z$ from the MUT and randomly selects $n$ \textit{pre-computed latent representations directly from $D_{val}$} to form $R(D_{com})$. Finally, NTD quantifies the similarity between $R(x)$ and $R(D_{com})$ and compares the quantified similarity with the pre-determined threshold (validated in Section~\ref{sec:animal}).

\mypara{Optimizing Online FRR and FAR}
NTD by default allows only one trial for an input. If the input from a user is (falsely) rejected, then it cannot appear again, which can be relaxed in some real-world scenarios, e.g., a face recognition application allows  multiple trials for a user to authenticate herself. 
We can leverage this relaxation to further improve NTD's detection usability (i.e., online FRR) and effectiveness (i.e., online FAR). 

Specifically, we preset relatively large FRR in the offline phase so that offline FAR is set to be small when deciding the threshold. 
For the online detection, both ${\rm FRR}_m$ and ${\rm FAR}_m$ of $m$ trials can be derived from offline FRR and FAR below:
\begin{equation}\label{eq:FRRmFARm}
\begin{cases}
      {\rm FRR}_m = {\rm FRR}^m\\
      {\rm FAR}_m = 1-(1-{\rm FAR})^m
\end{cases}
\end{equation}
Since we preset offline FRR that is larger than that of the only-one trial, ${\rm FRR}_m$ decreases significantly under $m$ trials and becomes much smaller than that of the only-one trial. Also, smaller offline FAR indicates smaller ${\rm FAR}_m$ (validated in Section~\ref{sec:animal}).

\subsubsection{Quantifying the Similarity}\label{sec:simMetric}

To quantify the similarity, we consider three different similarity metrics, i.e., cosine similarity, pearson correlation, and Tanimoto coefficient. Each metric can be used independently and introduced below. 

\mypara{Cosine Similarity} It is a measure of similarity of two non-binary vectors. Cosine Similarity between two vectors is defined below:

\begin{equation}
\begin{aligned}
Cos(A, B) = \frac{A\cdot B}{\left \| A \right \|\left \| B \right \|} = \frac{\sum\limits_{i=1}^{n}A_{i}\times B_{i}}{\sqrt{\sum\limits_{i=1}^{n}(A_{i})^{2}}\sqrt{\sum\limits_{i=1}^{n}(B_{i})^{2}}}
\end{aligned}
\end{equation}

\mypara{Pearson Correlation} It measures how well two sets of data fit on a straight line. Pearson Correlation between two vectors is defined below:

\begin{equation}
\begin{aligned}
Corr(A, B) = \frac{\sum\limits_{i=1}^{n}\left (A_{i} - \bar{A}\right)\times \left (B_{i} - \bar{B}\right)}{\sqrt{\sum\limits_{i=1}^{n} \left (A_{i} - \bar{A}\right)^{2}}\sqrt{\sum\limits_{i=1}^{n} \left (B_{i} - \bar{B}\right)^{2}}}
\end{aligned}
\end{equation}

\mypara{Tanimoto Coefficient} It handles the similarity of document data in text mining. Tanimoto Coefficient between two vectors is defined below: 

\begin{equation}
\begin{aligned}
T(A, B) 
= \frac{\sum\limits_{i=1}^{n}A_{i}\times B_{i}}{\sqrt{\sum\limits_{i=1}^{n}(A_{i})^{2}} + \sqrt{\sum\limits_{i=1}^{n}(B_{i})^{2}} - {\sum\limits_{i=1}^{n}A_{i}\times B_{i}}}
\end{aligned}
\end{equation}

We note that for each of the three similarity metrics above, $\textbf{A}$ and $\textbf{B}$ are the two vectors with length $\textbf{n}$. $\bar{\textbf{A}}$ is an average of all elements in \textbf{A}.

\eat{
\subsubsection{Metrics in Quantifying Detection Effectiveness}\label{sec:detectionmetric}

The detection effectiveness is quantitatively measured by false rejection rate (FRR) and false acceptance rate (FAR).

\vspace{2pt}\noindent$\bullet$ \textbf{FRR }is the probability that a benign input is regarded as a trigger input by NTD and thus falsely rejected.

\vspace{2pt}\noindent$\bullet$ \textbf{FAR }is the probability that the trigger input is regarded as a benign input by NTD and thus falsely accepted.
  
In practice, FRR represents the robustness of detection that affects the usability---a system should not frequently reject benign inputs, while FAR introduces security issues as it allows trigger inputs to pass. FRR and FAR can be balanced according to a specific application. 
}

\section{Evaluation}\label{sec:evaluation}
As NTD does not require any costly computing resources, we evaluate NTD on a common personal laptop, i.e., LENOVO with NVIDIA GeForce RTX 960M GPU, Intel i5-6300H CPU and 8GB DRAM memory. 

\subsection{Experimental Setup}
Before starting the evaluation, we introduce the three tasks and their reserved corresponding feature extractors as well as triggers.

\begin{table}[htbp]
    \centering
    \caption{A summary of datasets and their corresponding FEs.}
    \scalebox{0.80}{
    \begin{tabular}{c||c||c||c||c}
    \toprule
    \multirow{2}{*}{\begin{tabular}[c]{@{}c@{}}Task\\ Under Test\end{tabular}} & \multirow{2}{*}{\begin{tabular}[c]{@{}c@{}}FE\\Model Arch.\end{tabular}} & \multirow{2}{*}{\begin{tabular}[c]{@{}c@{}}\# of\\ Labels\end{tabular}} & \multicolumn{2}{c}{\begin{tabular}[c]{@{}c@{}}\# of Images\end{tabular}} \\ \cline{4-5} 
     &  &  & \begin{tabular}[c]{@{}c@{}}Used for\\ Threshold\end{tabular} & \begin{tabular}[c]{@{}c@{}}Used for\\ Detection\end{tabular} \\ \hline
    FaceScrub & FaceNet & 79 & 2,173 & 4,510 \\ \hline
    CTSDB & \begin{tabular}[c]{@{}c@{}}3Conv+\\ 2FC+1STN\end{tabular} & 57 & 1,838 & 4,324 \\ \hline
    Animals\_10 & ResNet50 & 10 & 7,848 & 18,331 \\ \bottomrule
    \end{tabular}}
    \label{tab:reservedFE}
\end{table}

\subsubsection{Tasks Under Test}
As shown in Table~\ref{tab:reservedFE}, we customize three datasets (i.e., FaceScrub~\cite{ng2014data}, Chinese Traffic Sign Database (CTSDB)~\cite{CTSDB}, and Animals\_10~\cite{Animals10}) for three distinct tasks. 
We use TensorFlow 2.2 to perform experiments with FaceScrub and Animals\_10 and Pytorch 1.8 with Python 3.7.9 to test CTSDB. 

\mypara{FaceScrub}
It is a public celebrity face dataset,including 106,863 facial images of 530 celebrities of men and women (about 200 images per person). It is currently one of the largest public face datasets and most of the images are with a size of $150 \times 150 \times 3$. We resize all of them to $160 \times 160 \times 3$ that is the FE accepted input size. We rely on~\cite{Overlap} to select 79 people whose identities \textit{do not overlap} with MS-Celeb-1M (another dataset used for the reserved feature extractor) and thus obtain 6,683 images in total.

\mypara{CTSDB} 
It has 58 categories, including 4,170 images for training and 1,994 images for the test. As our experiments do not need training, we merge both into one dataset for evaluation. The image size in this dataset ranges from $26 \times 28 \times 3$ to $200 \times 200 \times 3$. We resize all the images to $32 \times 32 \times 3$ that is the FE input size.
We note that category 009 (i.e., no straight through and right turn) from the dataset has only two samples, which is removed or unused. Thus, we have 57 categories with 6,162 samples in total. 

\mypara{Animals\_10} 
It is one customized dataset from Kaggle and contains 28K medium-quality collected from Google Images.
The animal images have 10 categories: dog, cat, horse, spider, butterfly, chicken, sheep, cattle, squirrel, and elephant. We resize all of them to $224 \times 224 \times 3$ that is the FE input size. 

For each dataset above, we divide it into two parts: offline dataset and online dataset.
The offline dataset serves as the held-out validation dataset $D_{val}$ for NTD, which will be used to determine the threshold. The online dataset is utilized to evaluate NTD's detection. 
Specifically, we randomly select 1000 samples from each online dataset as clean inputs and pass each to NTD for computing online FRR. To compute online FAR, we randomly select 1000 samples also from the online dataset and stamp them with a specified trigger as trigger inputs (see Section~\ref{sec:triggers}).
As $D_{val}$ should be small, the proportion of offline part to online part for each dataset is about 3:7 (see Table~\ref{tab:reservedFE}). When we evaluate a per-class threshold for each targeted class, we retain at least 11 samples for the class of the offline dataset (see details in Section~\ref{sec:faceresult}).

Before deploying NTD, we allow the maximized attacking success rate of the backdoor attack against each task (i.e., close to 100\% which is usually the case~\cite{gao2019strip,wang2019neural,guo2019tabor}), that is, trigger inputs are always predicted to be targeted classes/labels while clean inputs are correctly classified (e.g., the respective classification accuracy for the SOTA face recognition and traffic sign are 99.85\%~\cite{faceSOTA} and 99.71\%~\cite{trafficSOTA}.)

\subsubsection{Reserved Feature Extractors}
For each task, we reserve a corresponding FE, shown in Table~\ref{tab:reservedFE}.For FaceScrub, the FE~\cite{PretrainWieghtFace} is reserved from a public model (i.e., FaceNet~\cite{schroff2015facenet}), which is trained on the MS-Celeb-1M dataset~\cite{guo2016ms}. MS-Celeb-1M is generated by two steps. First, 100,000 from 1 million celebrities are selected based on their popularity. Then, about 100 samples for each selected celebrity are searched from a search engine. Thus, this dataset has 10 million images and each image is labeled by Microsoft.

For CTSDB, the FE~\cite{PretrainWieghtSign} is reserved {from} a public model that has 3 convolution layers (3Conv), 2 fully connected layers (2FC) and 1 spatial transformer network layer (1STN)~\cite{jaderberg2015spatial}. The model is trained on the German Traffic Sign Benchmark (GTSRB) dataset~\cite{Stallkamp2012}. GTSRB has more than 40 categories and more than 50,000 traffic signs. 

For Animals\_10, the FE~\cite{PretrainWieghtImagenet} is reserved {from} a public model (i.e., ResNet50~\cite{he2016deep}), which is trained on ImageNet dataset~\cite{deng2009imagenet}. ImageNet is a large-scale tagged image dataset, with 22,000 categories and 15 million images (each image has been tagged by rigorous manual screening).

Note that each FE above is available from a public platform for free, which is simply leveraged by NTD \textit{without} additional re-training or fine-tuning. As mentioned before, the FaceScrub has no overlap with the MS-Celeb-1M task.

\subsubsection{Triggers}\label{sec:triggers}
\begin{figure}[htb]
    \centering
    \includegraphics[width=0.48\textwidth]{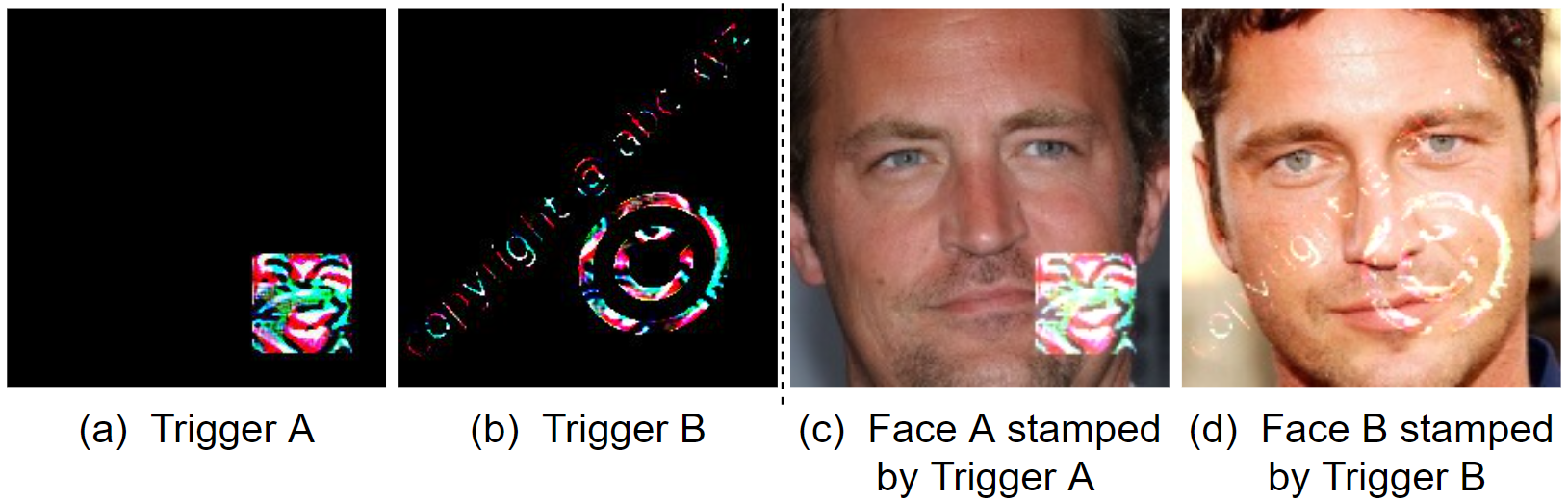}
    \caption{(left) Two triggers are used in our evaluation. (right) Face image samples are stamped by the triggers.}
    \label{fig:trigger}
\end{figure}

Without loss of generality, we select two popular triggers used by previous works~\cite{wang2019neural,guo2020trojannet}. For example, the selected triggers and their corresponding trigger inputs of the FaceScrub task are shown in Figure~\ref{fig:trigger}. {For each task, we apply Trigger A by default unless otherwise stated.}

\subsection{Face Recognition Classification}\label{sec:faceresult}

For this task, we consider four experimental settings below:

\mypara{Per-Class Threshold}
In this setting, we evaluate NTD using per-class threshold and thus {randomly} select three persons as targeted classes, i.e., `Gerard Butler', `Caroline Dhavernas' and `Edie Falco'.To decide the per-class threshold for each person, we set the size of the comparison set (i.e., $n$) to 3, choose the pearson correlation metric and preset different FRRs in the offline phase. 

\begin{figure}[htb]
    \centering
    \includegraphics[width=0.48\textwidth]{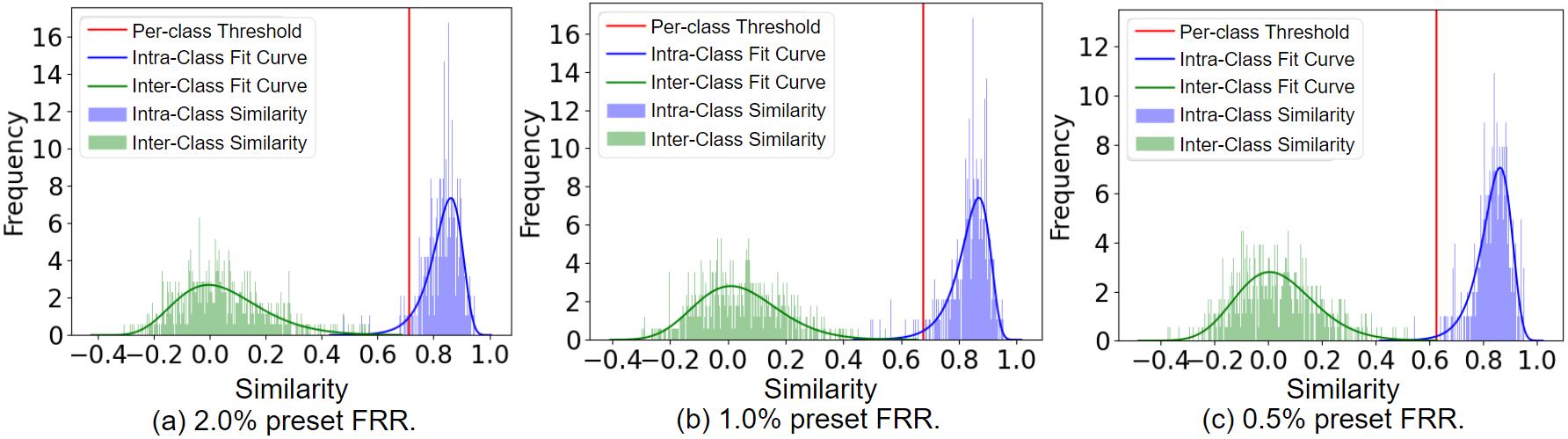}
    \caption{The online similarity distributions for both intra-class and inter-class, showing that there is almost no overlap between the two distributions with different offline FRRs preset.}
    \label{fig:face_frroffline_result}
\end{figure}

We display NTD's online detection results in Table~\ref{tab:facePerClass}. For each person, online FAR is as low as almost 0\% and insensitive to preset FRRs, indicating that NTD has effectively detected trigger inputs for this task. This is because that there is no intersection between the intra-class and the inter-class similarity distributions  during online detection. For example, Figure~\ref{fig:face_frroffline_result} shows the online similarity distribution for a targeted person, i.e., `Gerard Butler'. We have a similar offline similarity distribution for both intra-class and inter-class presents a similar curve when deciding a per-class threshold. We note that  the slight variance between the preset offline FRR and the online FRR, which is probably due to the relatively small number of testing rounds (i.e., 1000).

\begin{table}[]
    \centering
    \caption{NTD's detection results for FaceScrub using per-class thresholds.}
    \scalebox{0.80}{
    \begin{tabular}{c||c||c||c||c}
    \toprule
    Person & \begin{tabular}[c]{@{}c@{}}Preset\\ FRR (\%)\end{tabular} & Threshold & \begin{tabular}[c]{@{}c@{}}Online \\ FRR (\%)\end{tabular} & \begin{tabular}[c]{@{}c@{}}Online \\ FAR (\%)\end{tabular} \\ \hline
    \multirow{3}{*}{\begin{tabular}[c]{@{}c@{}}Gerard \\  Butler\end{tabular}} & 2.0 & 0.71321 & 5.1 & 0.0 \\ \cline{2-5} 
     & 1.0 & 0.67643 & 2.3 & 0.0 \\ \cline{2-5} 
     & 0.5 & 0.62588 & 1.2 & 0.0 \\ \hline
    \multirow{3}{*}{\begin{tabular}[c]{@{}c@{}}Caroline \\ Dhavernas \end{tabular}} & 2.0 & 0.52418 & 4.7 & 0.0 \\ \cline{2-5} 
     & 1.0 & 0.48235 & 2.8 & 0.0 \\ \cline{2-5} 
     & 0.5 & 0.43104 & 1.9 & 0.3 \\ \hline
    \multirow{3}{*}{\begin{tabular}[c]{@{}c@{}}Edie \\ Falco \end{tabular}} & 2.0 & 0.65735 & 1.4 & 0.0 \\ \cline{2-5} 
     & 1.0 & 0.63518 & 0.5 & 0.0 \\ \cline{2-5} 
     & 0.5 & 0.59298 & 0.0 & 0.1 \\ \bottomrule
    \end{tabular}}
    \label{tab:facePerClass}
\end{table}

\begin{table}[htbp]
    \centering
    \caption{NTD's detection effectiveness for FaceScrub is insensitive to similarity metrics, that is, all online FARs are almost 0\% for different metrics.}
    \scalebox{0.80}{
    \begin{tabular}{c||c||c||c||c}
    \toprule
    Person & \begin{tabular}[c]{@{}c@{}}Similarity\\ Metric\end{tabular} & Threshold & \begin{tabular}[c]{@{}c@{}}Online \\ FRR (\%)\end{tabular} & \begin{tabular}[c]{@{}c@{}}Online \\ FAR (\%)\end{tabular} \\ \hline
    \multirow{3}{*}{\begin{tabular}[c]{@{}c@{}}Matthew \\ Perry \end{tabular}} & Cosine & 0.45960 & 0.9 & 0.1 \\ \cline{2-5} 
     & Pearson & 0.51077 & 1.6 & 0.0 \\ \cline{2-5} 
     & Tanimoto & 0.33502 & 1.8 & 0.0 \\ \hline
    \multirow{3}{*}{\begin{tabular}[c]{@{}c@{}}Carey \\ Lowell \end{tabular}} & Cosine & 0.52969 & 1.4 & 0.0 \\ \cline{2-5} 
     & Pearson & 0.53659 & 1.4 & 0.0 \\ \cline{2-5} 
     & Tanimoto & 0.35744 & 1.0 & 0.0 \\ \hline
     \multirow{3}{*}{\begin{tabular}[c]{@{}c@{}}Kristin \\ Chenoweth \end{tabular}} & Cosine & 0.46144 & 0.9 & 0.1 \\ \cline{2-5} 
     & Pearson & 0.47296 & 0.9 & 0.0 \\ \cline{2-5} 
     & Tanimoto & 0.28683 & 0.5 & 0.0 \\ \bottomrule
    \end{tabular}}
    \label{tab:similarityMetric}
\end{table}
\mypara{Similarity Metrics}
In this setting, we evaluate three aforementioned similarity metrics mentioned in Section~\ref{sec:simMetric} to check whether NTD's detection is sensitive to a similarity metric. Thus, we replace the pearson correlation with the cosine similarity and the tanimoto coefficient, respectively, and perform similar experiments as in the previous setting. 

\begin{figure}[htb]
    \centering
    \includegraphics[width=0.48\textwidth]{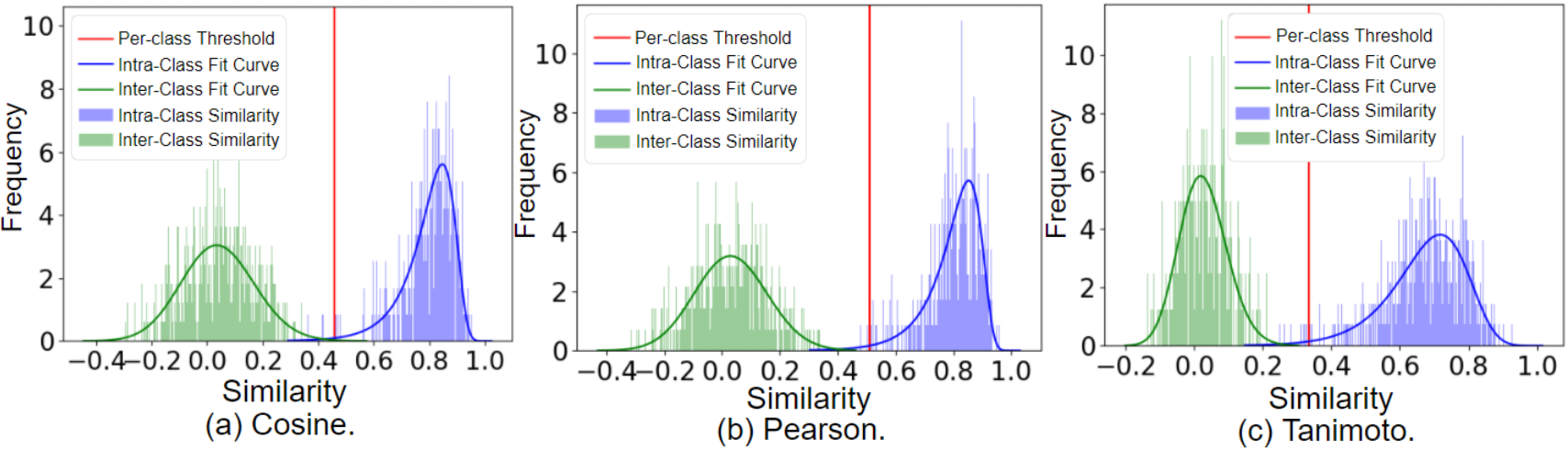}
    \caption{The online similarity distributions for both intra-class and inter-class, indicating that there is nearly no overlap between the two distributions in different similarity metrics.}
    \label{fig:face_metric_result}
\end{figure}

Specifically, we {randomly} select three persons, set $n$ to 3 and offline FRR to 0.5\% to determine per-class thresholds in the offline phase. 
The detection results displayed in Table~\ref{tab:similarityMetric} show that 
the online FARs are almost 0\% for all the selected similarity metrics, indicating that NTD's effectiveness is insensitive to the metrics.
Also, there is no overlap between the intra-class similarity distribution and the inter-class similarity distribution. As an example, we show  the online similarity distributions for a targeted person (i.e., `Matthew Perry') in Figure~\ref{fig:face_metric_result} with 0.5\% preset offline FRR and $n=3$.
In the following two settings, we choose the tanimoto coefficient without loss of generality.

\begin{table}[htbp]
    \centering
    \caption{NTD's detection results for FaceScrub using different comparison set sizes.}
    \scalebox{0.80}{
    \begin{tabular}{c||c||c||c||c}
    \toprule
    Person & \begin{tabular}[c]{@{}c@{}}Comparison \\Set Size \end{tabular} & Threshold & \begin{tabular}[c]{@{}c@{}}Online \\ FRR (\%)\end{tabular} & \begin{tabular}[c]{@{}c@{}}Online \\ FAR(\%)\end{tabular} \\ \hline
    \multirow{3}{*}{\begin{tabular}[c]{@{}c@{}}Gillian \\ Anderson \end{tabular}} & 5 & 0.32756 & 1.8 & 0.0 \\ \cline{2-5} 
     & 4 & 0.29166 & 0.6 & 0.1 \\ \cline{2-5} 
     & 3 & 0.29986 & 0.5 & 0.0 \\ \hline
    \multirow{3}{*}{\begin{tabular}[c]{@{}c@{}}Ken \\ Watanabe \end{tabular}} & 5 & 0.54259 & 1.9 & 0.0 \\ \cline{2-5} 
     & 4 & 0.53706 & 1.0 & 0.0 \\ \cline{2-5} 
     & 3 & 0.52026 & 1.5 & 0.0 \\ \hline
     \multirow{3}{*}{\begin{tabular}[c]{@{}c@{}}Woody \\ Allen\end{tabular}} & 5 & 0.36479 & 0.0 & 0.1 \\ \cline{2-5} 
     & 4 & 0.34259 & 0.0 & 0.1 \\ \cline{2-5} 
     & 3 & 0.32593 & 0.0 & 0.1 \\ \bottomrule
    \end{tabular}}
    \label{tab:groupsize}
\end{table}

\begin{figure}[htb]
    \centering
    \includegraphics[width=0.48\textwidth]{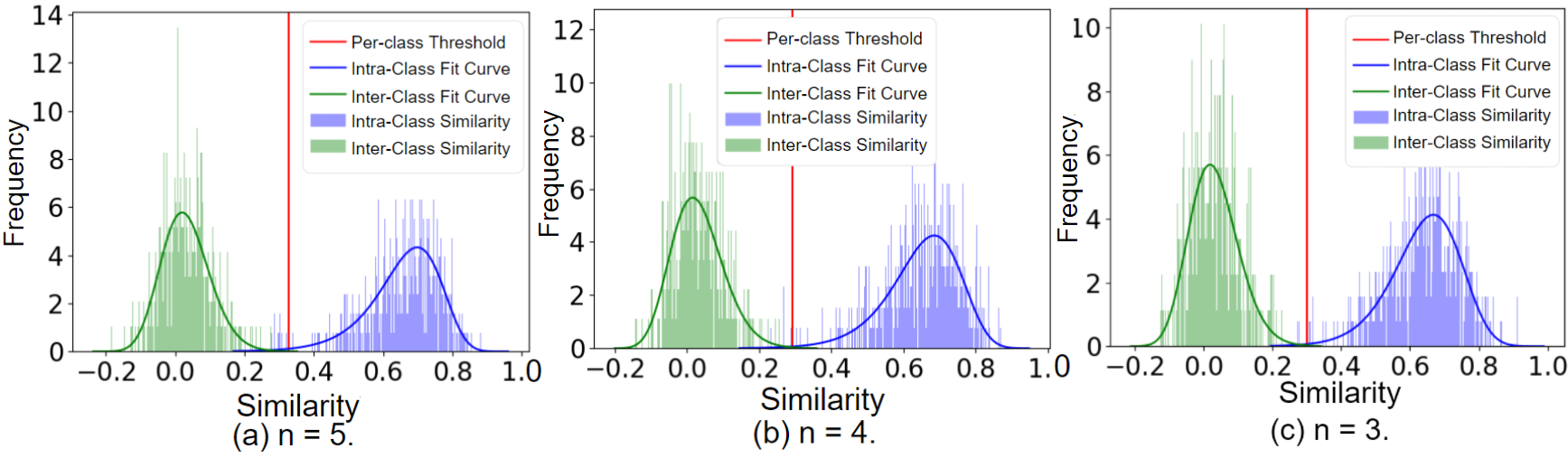}
    \caption{The online similarity distributions for both intra-class and inter-class, presenting virtually no overlap between the two distributions in different comparison set sizes (i.e., $n$).}
    \label{fig:face_groupsize_result}
\end{figure}

\mypara{Comparison Set Size} 
For previous settings, we use a small size of the comparison set, i.e., $n = 3$.
In this setting, we increase $n$ to check whether a larger size will increase the detection effectiveness. If not, we can keep $n$ to 3 to retain a low latency as well as manage a small validation set.
Specifically, we set $n$ to 3, 4, and 5, respectively. For each $n$, we preset offline FRR to 0.5\%, use the tanimoto coefficient as the metric and perform similar experiments for three randomly chosen person classes. 
We display NTD's detection results in Table~\ref{tab:groupsize}. For each person under different $n$, the online FARs are almost 0\%, indicating that NTD's detection effectiveness is insensitive to the comparison set size and there exists virtually no overlap in the similarity distributions between intra-class and inter-class.
Take Figure~\ref{fig:face_groupsize_result} as an example, it shows the online similarity distributions for a targeted person (i.e., ‘Gillian  Anderson’) with different comparison set sizes.

\mypara{Global Threshold}
In certain scenarios where all classes have similar security sensitivity or attacker-targeted classes have no enough validation samples to decide per-class thresholds, we need to decide a global threshold for all classes. Thus, we set $n$ to 3, use the tanimoto coefficient as a similarity metric, randomly select 1000 different samples (i.e., $x$) from all classes, preset offline FRR to obtain a global threshold.

\begin{figure}[htb]
    \centering
    \includegraphics[width=0.48\textwidth]{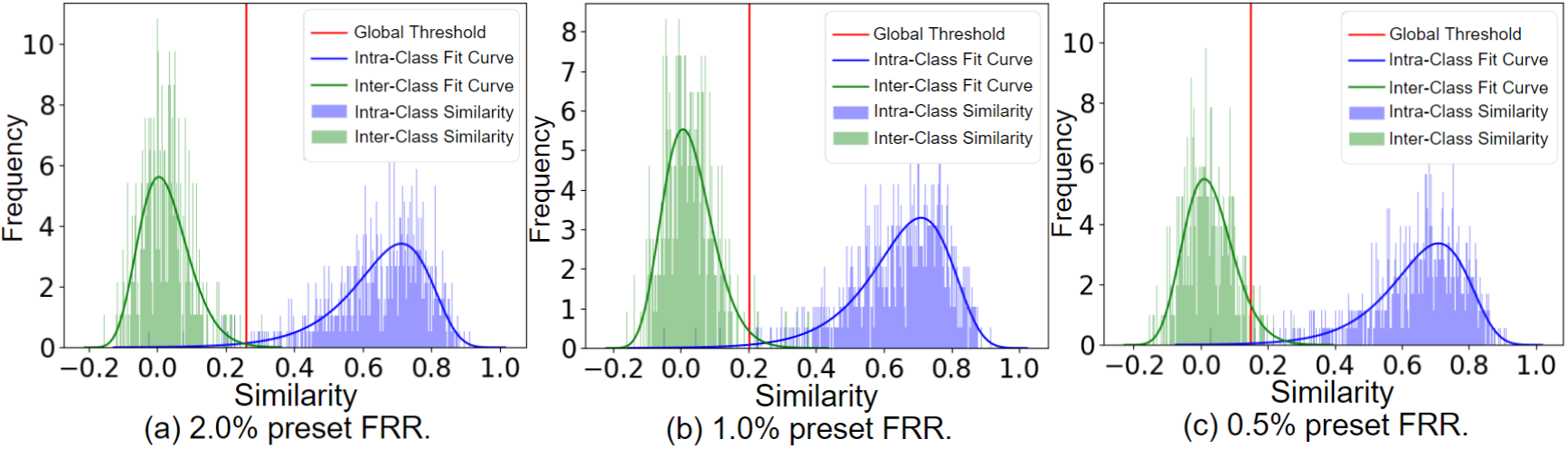}
    \caption{The online similarity distributions for both intra-class and inter-class, showing that there exists an overlap between the two distributions with different offline FRRs preset.}
    \label{fig:global_result}
\end{figure}

Table~\ref{tab:global} shows the detection results using global thresholds under different (offline) preset FRRs. 
Clearly, the online detection is usable as all the online FRRs for both triggers are within 2.3\%. 
As NTD in the setting of per-class threshold is effective with different present FRRs (i.e., online FARs are all about 0\%), 
NTD's effectiveness increases (online FARs decreases) when present FRRs increases.
This is because that there exists an overlap between the intra-class similarity distribution and the inter-class similarity distribution, displayed in Figure~\ref{fig:global_result}. 
Particularly, the global threshold becomes larger when the preset FRR increases, thus making NTD better detect trigger inputs (i.e., lower online FARs). 
The root cause is that we obtain the intra-class similarity distribution for 1000 random samples that can come from different classes. In the setting of per-class threshold, the intra-class similarity distribution is computed based on 1000 random samples from \textit{the same class}, thus resulting in a smaller variance and better detection effectiveness.

We also decide a global threshold by selecting samples coming from specified classes rather than all the classes, i.e., 40 specified persons out of the whole 79 persons. 
The detection results are comparable with the results in Table~\ref{tab:global}, indicating that it may be enough to use a certain number of classes when deciding a global threshold. When the number of classes for a task is large, this method can save time for deciding a global threshold in the offline phase.

\begin{table}[]
    \centering
    \caption{NTD's detection results for FaceScrub using global thresholds.}
    \scalebox{0.70}{
    \begin{tabular}{c||c||c||c||c||c}
    \toprule
    Person & Trigger & \begin{tabular}[c]{@{}c@{}}Preset\\ FRR (\%)\end{tabular} & Threshold & \begin{tabular}[c]{@{}c@{}}Online\\ FRR (\%)\end{tabular} & \begin{tabular}[c]{@{}c@{}}Online \\ FAR (\%)\end{tabular} \\ \hline
    \multirow{6}{*}{Global} & \multirow{3}{*}{A} & 0.5 & 0.15085 & 0.1 & 5.8 \\ \cline{3-6} 
     &  & 1.0 & 0.20439 & 0.7 & 1.9 \\ \cline{3-6} 
     &  & 2.0 & 0.25970 & 0.9 & 0.5 \\ \cline{2-6} 
     & \multirow{3}{*}{B} & 0.5 & 0.10588 & 0.2 & 15.8 \\ \cline{3-6} 
     &  & 1.0 & 0.20015 & 0.9 & 2.2 \\ \cline{3-6} 
     &  & 2.0 & 0.32886 & 2.3 & 0.2 \\  \bottomrule
    \end{tabular}}
    \label{tab:global}
\end{table}

\observ{(1) Per-class threshold is more effective in online detection than global threshold. 
(2) The similarity metric and trigger size have little effect on  NTD.
(3) The comparison set size only affects NTD's detection latency, that is,  smaller size achieves lower detection latency}

\subsection{Traffic Sign Classification}

\begin{table}[htb]
    \caption{NTD's detection results for CTSDB using global thresholds (5\% FRR preset).}
    \scalebox{0.76}{
    \begin{tabular}{c||c||c||c||c||c}
    \toprule
    Sign & Trigger & \begin{tabular}[c]{@{}c@{}}Comparison Set \\ Size \end{tabular} & Threshold & \begin{tabular}[c]{@{}c@{}}Online \\ FRR (\%)\end{tabular} & \begin{tabular}[c]{@{}c@{}}Online \\FAR (\%)\end{tabular} \\ \hline
    \multirow{6}{*}{Global} & \multirow{3}{*}{A} & 3 & 0.50461 & 3.9 & 3.5 \\ \cline{3-6} 
     &  & 5 & 0.53791 & 4.7 & 4.4 \\ \cline{3-6} 
     &  & 7 & 0.52995 & 4.7 & 2.2 \\ \cline{2-6} 
     & \multirow{3}{*}{B} & 3 & 0.50796 & 4.1 & 2.5 \\ \cline{3-6} 
     &  & 5 & 0.53061 & 4.6 & 2.2 \\ \cline{3-6} 
     &  & 7 & 0.54969 & 5.3 & 1.6 \\ \bottomrule
    \end{tabular}}
    \label{tab:sign_global_result}
\end{table}

\begin{table}[htb]
    \centering
    \caption{NTD's detection results of individual categories for CTSDB using a global threshold corresponding to 5\% preset FRR and the comparison set size being 3.
    % , threshold = 0.50461 (CTSDB).
    }
    \scalebox{0.8}{
    \begin{tabular}{c||c||c}
    \toprule
    Sign & \begin{tabular}[c]{@{}c@{}}Online FRR (\%)\end{tabular} & \begin{tabular}[c]{@{}c@{}}Online FAR (\%)\end{tabular} \\ \hline
    002 & 0.0 & 9.0 \\ \hline
    012 & 6.6 & 3.9 \\ \hline
    024 & 0.1 & 0.4 \\ \hline \hline
    037 & 22.4 & 8.9 \\ \hline
    043 & 9.2 & 4.4 \\ \hline
   050 & 0.0 & 11.7 \\ \bottomrule
    \end{tabular}}
    \label{tab:sign_single_result}
\end{table}

\begin{table}[htb]
    \centering
    \caption{NTD's detection results for CTSDB using per-class-thresholds with the comparison set size being 3.}
    \scalebox{0.8}{
    \begin{tabular}{c||c||c||c||c}
    \toprule
    Sign & \begin{tabular}[c]{@{}c@{}}Preset \\ FRR (\%)\end{tabular} & Threshold & \begin{tabular}[c]{@{}c@{}}Online \\ FRR (\%)\end{tabular} & \begin{tabular}[c]{@{}c@{}}Online \\ FAR (\%)\end{tabular} \\ \hline
    \multirow{3}{*}{002} & 5 & 0.81645 & 11.5 & 0.0 \\ \cline{2-5} 
     & 3 & 0.80163 & 9.1 & 0.0 \\ \cline{2-5} 
     & 1 & 0.76246 & 2.3 & 0.0 \\ \hline
    010 & 5 & 0.55947 & 2.8 & 0.8 \\ \hline
    012 & 5 & 0.52046 & 7.3 & 4.1 \\ \hline
    024 & 5 & 0.58445 & 3.0 & 0.0 \\ \hline \hline
    037 & 5 & 0.41670 & 10.2 & 11.5 \\ \hline
    043 & 5 & 0.47755 & 6.6 & 7.1 \\ \hline
    050 & 5 & 0.58600 & 0.9 & 6.8 \\ \hline
    036 & 5 & 0.75778 & 3.4 & 0.4 \\ \hline
    \multirow{3}{*}{038} & 5 & 0.41813 & 5.8 & 12.7 \\ \cline{2-5} 
     & 7 & 0.44234 & 5.7 & 10.6 \\ \cline{2-5} 
     & 10 & 0.47799 & 7.3 & 7.4 \\ \hline
    \multirow{2}{*}{040} & 5 & 0.40589 & 2.5 & 11.0 \\ \cline{2-5} 
     & 7 & 0.42950 & 4.6 & 7.7 \\ \hline
    042 & 5 & 0.68266 & 7.6 & 1.0 \\ \hline
    044 & 5 & 0.60807 & 0.5 & 5.1 \\ \hline
    054 & 5 & 0.31364 & 6.9 & 4.0 \\ \bottomrule
    \end{tabular}}
    \label{tab:sign_ind_threshold_result}
\end{table}

\begin{table}[htb]
    \centering
    \caption{NTD's detection results for CTSDB using per-class-thresholds (the comparison set size being 3) when more noisy background features of the images are removed.}
    \scalebox{0.80}{
    \begin{tabular}{c||c||c||c||c}
    \toprule
    Sign & \begin{tabular}[c]{@{}c@{}}Preset \\ FRR (\%)\end{tabular} & Threshold & \begin{tabular}[c]{@{}c@{}}Online \\ FRR (\%)\end{tabular} & \begin{tabular}[c]{@{}c@{}}Online \\ FAR  (\%)\end{tabular} \\ \hline
    010 & 5 & 0.57054 & 3.5 & 0.9 \\ \hline
    012 & 5 & 0.19765 & 5.1 & 29.3 \\ \hline \hline
    043 & 5 & 0.51654 & 5.7 & 2.9 \\ \hline
    040 & 7 & 0.43530 & 4.2 & 6.3 \\ \bottomrule
    \end{tabular}}
    \label{tab:sign_resize_result}
\end{table}

\begin{figure}[htb]
    \centering
    \includegraphics[width=0.40\textwidth]{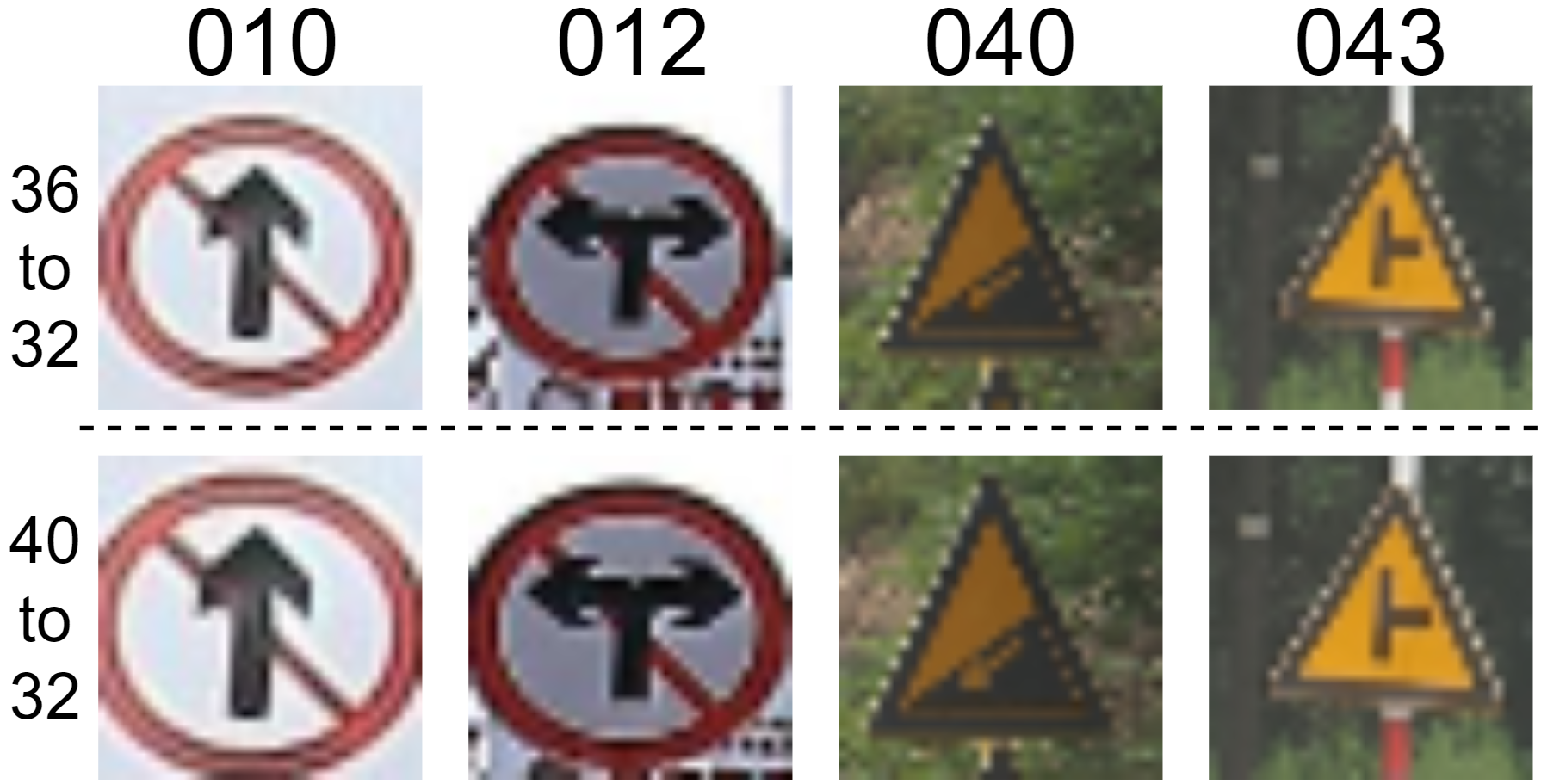}
    \caption{The traffic sign images are preprocessed to reduce background noise.}
    \label{fig:centercrop}
\end{figure}
For this task, we consider two experimental settings for evaluation, i.e., global threshold and per-class threshold, and use
the pearson correlation as the similarity metric. Before we start, we need to preprocess the traffic sign images of the task as they have much background noise. Specifically, we resize the images to $36 \times 36 \times 3$ and then crop the center area of the images to be $32 \times 32 \times 3$, as shown at the top of Figure~\ref{fig:centercrop}.

\mypara{Global Threshold}
With 5\% offline FRR preset, we decide global thresholds under different comparison set sizes. After that, NTD uses the thresholds to check clean inputs or trigger inputs with either Trigger A or B, results of which are shown in Table~\ref{tab:sign_global_result}. 
Both online FARs and FRRs are around 5\% for either trigger, indicating that NTD is effective in detecting trigger inputs and usable for benign inputs. To verify whether NTD is effective for each category of the task by using the pre-determined thresholds, we select 6 typical categories (with Trigger A applied) and show them in Table \ref{tab:sign_single_result}. Clearly, NTD has different online FRRs and FARs for different categories, and some categories (e.g., 024) are with better results than the global-threshold determined results in  Table~\ref{tab:sign_global_result} while some are not (e.g., 037).

\mypara{Per-Class Threshold} 
In this setting, we evaluate NTD using per-class-thresholds for more than 20 selected categories that can be attacker-targeted using Trigger A and present the results in Table~\ref{tab:sign_ind_threshold_result}. We divide the selected categories into two parts based on whether 
features of a given category are similar to that of a category from GTSRB that trained the corresponding FE. 

For the part of categories that have similar main features, NTD is effective as expected in detecting trigger inputs targeting these categories and we show four of them at the top of Table~\ref{tab:sign_ind_threshold_result} (i.e., 002, 010, 012 and 024)

For the other part of categories with less similar main features, all of them are shown at the bottom of Table~\ref{tab:sign_ind_threshold_result}. 
The majority of the categories in this part have lower online FARs compared to that with a global threshold as in Table~\ref{tab:sign_global_result}, indicating that a per-class threshold is fine-grained and works better than a coarse-grained global threshold for the CTSDB task.
We can further improve NTD for this task by removing more irrelevant and noisy background feature in the traffic sign image as detailed below.

\mypara{Noisy Background Features} Unlike previous preprocessing operations that resize the images to  $36 \times 36 \times 3$, we boldly resize the traffic sign images to $40 \times 40 \times 3$ before cropping the image into $32\times 32 \times 3$, as visualized at the bottom of Figure~\ref{fig:centercrop}. These resizing and cropping operations will zoom-in on the center area of an image and thus remove more noisy background. We then perform similar experiments for four categories as in the previous setting of per-class threshold and display the results in Table~\ref{tab:sign_resize_result}. 

For the categories at the top of Table~\ref{tab:sign_resize_result}, they have similar main features as mentioned above and the online FARs for them are unexpected to increase. This is probably because that there are just a few noisy background features for these categories and the main traffic sign features are removed,  therefore, there is no need to perform further resizing and cropping operations for these categories. 
For the categories at the bottom, the online FARs for both 043 and 040 have reduced greatly compared with that in Table~\ref{tab:sign_ind_threshold_result}, though in different degrees. The latter probably because their noisy background features are different.

\observ{(1) NTD is insensitive to trigger size. (2) Compared to global threshold, per-class threshold enables better detection effectiveness. (3) Reducing noisy background features can help improve the detection effectiveness
}

\subsection{General Animal Classification}\label{sec:animal}
For this task, we consider three experimental settings. First, we apply a fine-grained per-class threshold for each class as this task has only 10 classes. Second, we evaluate the impacts of multiple trials on NTD's detection. Last, we experiment with online detection latency. As this task has many samples for each category, we set the size of the comparison set (i.e., $n$) to 20. The pearson correlation is used as the similarity metric.

\begin{table}[]
    \centering
    \caption{NTD's detection results for Animals\_10 using per-class-thresholds.}
    \scalebox{0.753}{
    \begin{tabular}{c||c||c||c||c}
    \toprule
    Animal & \begin{tabular}[c]{@{}c@{}}Preset\\ FRR (\%)\end{tabular} & Threshold & \begin{tabular}[c]{@{}c@{}}Online \\ FRR (\%)\end{tabular} & \begin{tabular}[c]{@{}c@{}}Online \\ FAR (\%)\end{tabular} \\ \hline
    \multirow{2}{*}{elephant} & 6.0 & 0.35488 & 6.3 & 1.2 \\ \cline{2-5} 
     & 5.0 & 0.32066 & 4.2 & 2.6 \\ \hline
    chicken & 6.0 & 0.26021 & 5.8 & 3.9 \\ \hline
    \multirow{2}{*}{cat} & 6.0 & 0.28049 & 8.7 & 1.3 \\ \cline{2-5} 
     & 5.0 & 0.24817 & 3.9 & 4.4 \\ \hline
    \multirow{3}{*}{spider} & 6.0 & 0.29316 & 7.0 & 0.6 \\ \cline{2-5} 
     & 5.0 & 0.27084 & 5.1 & 2.0 \\ \cline{2-5} 
     & 4.0 & 0.26076 & 4.6 & 1.6 \\ \hline
    squirrel & 6.0 & 0.28303 & 5.0 & 2.3 \\ \hline \hline
    \multirow{2}{*}{butterfly} & 6.0 & 0.21997 & 7.4 & 8.7 \\ \cline{2-5} 
     & 7.0 & 0.23071 & 9.4 & 8.5 \\ \hline \hline
    \multirow{2}{*}{horse} & 6.0 & 0.31812 & 6.5 & 1.6 \\ \cline{2-5} 
     & 5.0 & 0.29251 & 3.5 & 3.4 \\ \hline
    \multirow{2}{*}{cattle} & 6.0 & 0.31689 & 8.8 & 1.3 \\ \cline{2-5} 
     & 5.0 & 0.29567 & 5.6 & 2.8 \\ \hline
    \multirow{2}{*}{sheep} & 6.0 & 0.24115 & 4.9 & 11.3 \\ \cline{2-5} 
     & 7.0 & 0.24808 & 7.3 & 6.2 \\ \hline \hline
    dog & 6.0 & 0.15772 & 6.7 & 23.5 \\ \toprule
    \end{tabular}}
    \label{tab:animal_result}
\end{table}

\eat{
\begin{figure}[htb]
    \centering
    \includegraphics[width=0.30\textwidth]{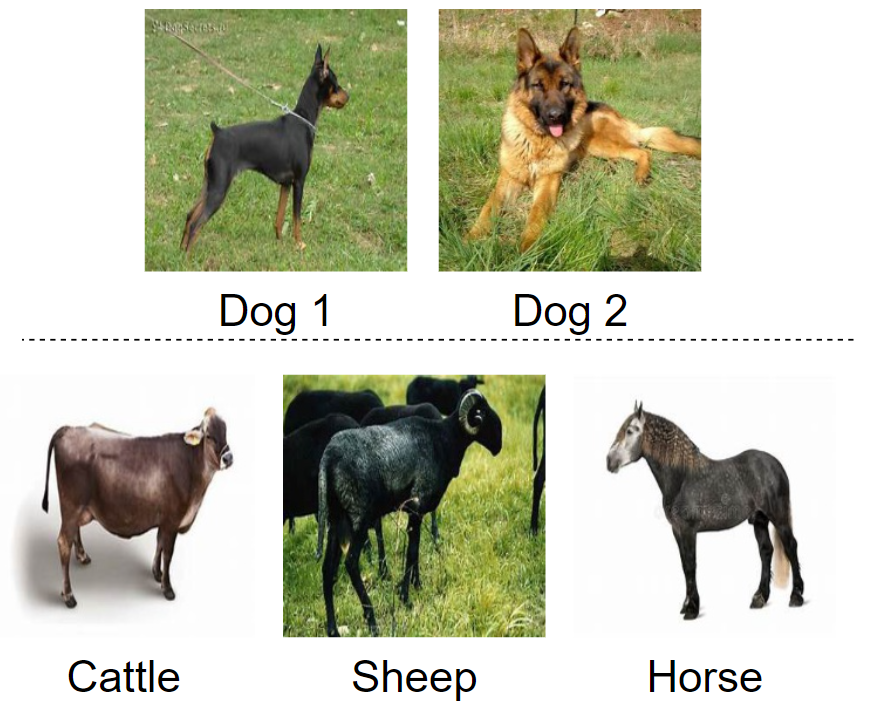}
    \caption{Examples of cattle, horse and sheep and dog. The first three animals are similar, which the dog also share certain similarities with them.}
    \label{fig:dog}
\end{figure}
}

\mypara{Per-Class Threshold}
As shown in Table~\ref{tab:animal_result}, we preset different offline FRRs for each class and decide a corresponding per-class threshold. For the detection results in the table, we discuss them in the following four aspects:

\begin{itemize}
    
\item As the first five categories (i.e., `elephant', `chicken', `cat', `spider', `squirrel') present high intra-class similarity and low inter-class similarity in the offline phase, both online FRRs and FARs are relatively low. Particularly, online FARs are within 4.4\%. If these categories are targeted by the attacker, NTD's detection effectiveness reaches more than 95\%.

\item As the `butterfly' category is with relatively low intra-class {and low inter-class similarity} in the offline phase, its per-class threshold does not distinguish inputs of this category well and we obtain a higher online FAR than the previous five categories. The root cause for the low intra-class similarity is the large differences among the samples in this category of the offline dataset.If this category is targeted, NTD is less effective but still reaches about 92\%.

\item As the four categories (i.e., `horse', `cattle', `sheep' and `dog') share certain features (e.g., all of them have four legs and relatively similar tails), they have relatively high inter-class similarities. To reduce the inter-class similarity for each of the four categories, the samples from each inter-class comparison set are all from the other six categories. Thus, NTD reaches more than 94\% effectivenss in detecting trigger inputs for the `horse', `cattle' and `sheep' categories. 

\item Besides the high inter-class similarity, the last `dog' category has relatively low intra-class similarity, making the intersection between the two similarity distributions not small and thus rendering NTD's detection not effective. When the attacker targets this category, NTD's detection effectiveness is reduced to about 76\%. This is the limitation of NTD, that is, NTD requires intra-class similarity to be higher than that of inter-class similarity (see more discussions in Section~\ref{sec:limitation}).

\end{itemize}

\mypara{Optimizing Online FRR and FAR}
To improve online FRR and FAR for all the categories, we can apply multiple trials where we select five animals for our experiment. For a per-class threshold of each class, we preset the offline FRR to 20\% and the number of trials (i.e., $m$) to be either 2 or 3. The results are detailed in Table~\ref{tab:multipleTrial}. When $m$ is 3, the empirical online ${\rm FAR}_m$ for each category is within 0.4\%, while the empirical online ${\rm FRR}_m$ also decreases (within 2.5\%), significantly improving both NTD's detection effectiveness and usability. Besides, the empirical ${\rm FRR}_m$ and ${\rm FAR}_m$ are comparable with the theoretical ones from Equation~\ref{eq:FRRmFARm}. Compared to Table~\ref{tab:animal_result}, the averaged online FRR/FAR have been reduced from 3.9\%/4.4\% with $m=1$ to 1.7\%/0.40\% with $m=3$.

\begin{table}[]
    \centering
    \caption{NTD's detection results for Animals\_10 when multiple trials (i.e., $m$) are available.}
    \scalebox{0.80}{
    \begin{tabular}{c||c||c||c||c}
    \toprule
    Animal & Threshold & $m$ & \begin{tabular}[c]{@{}c@{}}Online \\ ${\rm FRR}_m$(\%)$^*$ \end{tabular}  & \begin{tabular}[c]{@{}c@{}}Online \\ {${\rm FAR}_m$(\%)$^*$} \end{tabular} \\ \hline 
    \multirow{2}{*}{elephant} & \multirow{2}{*}{0.43915} & 2 & 6.20 ; 6.33 & 0.20 ; 0.20 \\ \cline{3-5} 
     & & 3 & 2.50 ; 1.57 & 0.20 ; 0.18 \\ \hline
    \multirow{2}{*}{chicken}  & \multirow{2}{*}{0.34629} & 2 & 5.00 ; 4.12 & 0.20 ; 0.20 \\ \cline{3-5} 
     & & 3 & 0.90 ; 0.97 & 0.20 ; 0.18 \\ \hline
    \multirow{2}{*}{cat}      & \multirow{2}{*}{0.35878} & 2 & 5.10 ; 4.82 & 0.30 ; 0.30 \\ \cline{3-5} 
     & & 3 & 1.70 ; 1.33 & 0.40 ; 0.39 \\ \hline
    \multirow{2}{*}{spider}   & \multirow{2}{*}{0.37746} & 2 & 6.60 ; 6.86 & 0.00 ; 0.00 \\ \cline{3-5} 
     & & 3 & 1.80 ; 1.62 & 0.00 ; 0.00 \\ \hline
    \multirow{2}{*}{squirrel} & \multirow{2}{*}{0.36991} & 2 & 5.90 ; 5.55 & 0.20 ; 0.20 \\ \cline{3-5} 
     & & 3 & 1.40 ; 0.98 & 0.30 ; 0.30 \\ \toprule
    \end{tabular}}
    \begin{tablenotes}
      \footnotesize
      \item $^*$: The format is $x;y$, where $x$ is obtained through our experiments and $y$ is computed from Equation~\ref{eq:FRRmFARm}.  
    \end{tablenotes}\label{tab:multipleTrial}		
\end{table}

\begin{figure}[htb]
    \centering
    \includegraphics[width=0.40\textwidth]{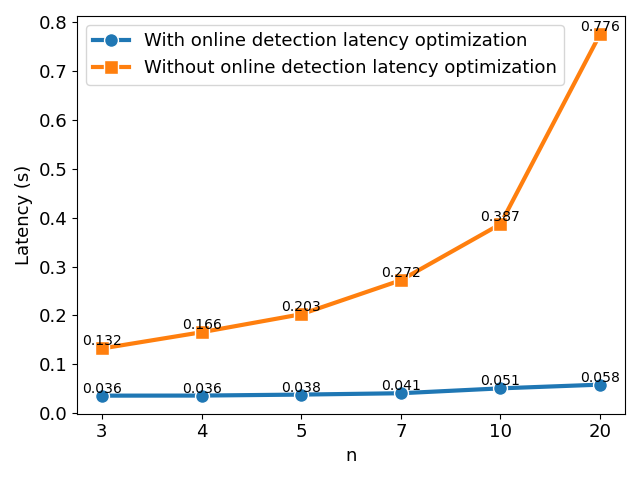}
    \caption{When the comparison set size becomes larger, NTD's latency per trial increases significantly  without online detection latency optimization. However, with the optimization, the latency is much lower and increases merely gradually.}
    \label{fig:timeLatency}
\end{figure}

\mypara{Optimizing Online Detection Latency}
As discussed in Section~\ref{sec:realtime}, the online detection latency can be optimized by retrieving features of the validation dataset from the FE in the offline phase. 
In this setting, we select 1000 images from the `cat' category as inputs $x$ to NTD and measure its detection latency under different comparison set size with/without the latency optimization, results of which are displayed in Figure~\ref{fig:timeLatency}.
The figure clearly shows that the latency optimization significantly reduces NTD's latency, especially when $n$ becomes larger. For example, the latency drops from 0.776s to 0.058s (i.e., $13\times$ reduction) when $n=20$, the default value for this task.

\observ{
(1) The online FRR and FAR optimization has greatly improved NTD's overall detection effectiveness and usability. (2) The online detection latency optimization has significantly reduced NTD's latency}

\section{Discussion}\label{sec:discussion}

\mypara{Fitting a Similarity Distribution} 
When deciding a threshold, we fit intra-class similarity distribution and inter-class similarity distributions using log-gamma curve and gamma curve, respectively.
We note that fitting the distribution is not a must. Alternatively, we can use a ranking method to determine the threshold. For example, if there are 1,000 intra-class similarity values, we rank them in an ascending manner. Given 5\% preset FRR, we simply pick the $50_{\rm th}$ smallest inter-class similarity value, and use it as the threshold. We have evaluated this method using the FaceScrub task and it demonstrates a similar detection effectiveness as the fitting approach. 

\mypara{Inspecting Offline Data Poisoning}
Some backdoor defenses~\cite{tran2018spectral,chen2018detecting,peri2020deep,tang2021demon} must inspect a poisoned offline dataset to mitigate backdoor attacks, which is not required by NTD. However, the user can indeed use NTD to detect trigger samples from the dataset when the entire training dataset is under access and perform rigorous human inspection against the detected samples to recover the trigger. 

\mypara{Inter-Class Similarity Over Intra-Class Similarity}\label{sec:limitation}
NTD is effective if intra-class similarity is higher than inter-class similarity. 
If a task or some classes within a task cannot meet the condition, NTD cannot be immediately adopted. As shown in the general animal classification, NTD is relatively ineffective in protecting the `dog' category while effective in protecting other categories. 
To check whether the condition can be met, the user simply leverages the reserved FE and the validation set to compare intra-class similarity with inter-class similarity even before training her task. Such pre-diagnosis is user-friendly to the user. 

\mypara{NTD's Detection on Other Tasks}
We focus on image-classification tasks to evaluate NTD and its effectiveness besides the image classification has not been investigated yet. Nonetheless, we believe that NTD can also work for other domain tasks such as audio and textal, especially classification tasks as long as the intra-class and inter-class similarity of a given task can be distinguished by a reserved FE, 

\mypara{Minimal Training on FE}
In this paper, NTD requires no ML training against a reserved FE. If an NTD user can train the FE, she can take lightweight fine-tuning on the FE to better extract the latent features from inputs for a model under test (MUT), which makes the FE more relevant to the MUT. 
To this end, she can leverage few-shot learning~\cite{sung2018learning} and siamese network~\cite{koch2015siamese} to assist the similarity check of intra-class and inter-class.

\section{Related Work}\label{sec:comparison}
Based on when a backdoor behavior is detected, existing countermeasures can be divided into two categories, i.e., offline~\cite{wang2019neural,tang2021demon,xu2019detecting} and online~\cite{liu2019abs,gao2019strip,doan2020februus} approaches. %According to which information is used, 
We can also categorize the backdoor countermeasures into data-level~\cite{tran2018spectral,tang2021demon,chen2018detecting,chou2020sentinet} and model-level~\cite{xu2019detecting,chen2019deepinspect,liu2019abs,wang2019neural,guo2019tabor} approaches. We describe state-of-the-art defenses and distinguish NTD from them. A qualitative comparison is summarized in Table~\ref{tab:defenseCompar}.

\mypara{Neural Cleanse~\cite{wang2019neural} (S\&P'19)} 
It is an offline model-based defense, which observes that there exists an extremely small perturbation that is applicable for all inputs to be misclasified into a victim label. And this perturbation resembles a real trigger. It iterates each label to identify a potential trigger. Among all ponytail triggers, the smallest one generally is chosen based on an outlier detection algorithm and its corresponding label is the attacker-targeted label. As it needs to go through each label, its computational overhead is dependent on the task, e.g., face recognition costs an extremely long time (up to several days). In addition, the performance of Neural Cleanse is contingent on trigger size, pattern, location and other factors~\cite{guo2019tabor}. In fact, in some cases, its detection is unstable, e.g., falsely rejecting a clean model or falsely accepting a backdoored model~\cite{guo2019tabor,tian2021stealthy,ma2021quantization} even when the attack is within its threat model. it is ineffective against source-specific backdoor attacks. The presented NTD eliminates these concerns.

\mypara{ABS~\cite{liu2019abs} (CCS'19)} 
ABS, short for Artificial Brain Stimulation, is an online model-based defense. It scans an ML model to find the target that significantly activates the neuron. It uses reverse engineering to distinguish between benign neurons and poisoned neurons. If there are more inputs that convert other labels into specific labels, then there is a high probability that the model is considered to be a backdoor attack model with triggers. However, this method assumes that the trigger only stimulates an internal neuron instead of a group of interacting neurons to increase activation. 
This assumption may not be true in advanced attacks, e.g., multiple patches spread across the whole image as a trigger. 
In addition, the ABS that is only effective against source-agnostic backdoor attack. The NTD has no these concerns.

\mypara{MNTD~\cite{xu2019detecting} (S\&P'21)} 
MNTD, short for Meta neural trojan detection, is an offline model-based defense. MNTD trains a meta classifier over a number of benign and backdoored ML models to infer whether an unforeseen model is backdoored or not. It requires training a large number of shadow models on the same task as the target model that will be checked. The MNTD has the advantage of defeating unknown attack strategies and generically being applicable to multiple domains such as vision, speech, and textual. However, it comes with the trade-off of heavy computational overhead. For example, it needs to train 4096 shadow models for the simple MNIST task, which takes about 14 hours. This appears to be magnitude computationally expensive comparing with training a clean model for the MNIST task by the users/defenders themselves. Training both the shadow models and the meta classifier do require ML expertise. Our presented NTD is also an attack strategy independent as the MNTD. Meanwhile, NTD voids the requirements of additional expensive training and ML expertise.

\mypara{SCAn~\cite{tang2021demon} (USENIX'21)} 
SCAn, short for Statistical contamination analyzer, is an offline data-based defense. Though SCAn can defeat the challenging source-specific backdoor attacks, it needs to access the training data that must contain the trigger samples to check and/or sanitize the data, which has a similar assumption to~\cite{tran2018spectral,chen2018detecting}. This assumption constrains its applicability~\cite{tran2018spectral,chen2018detecting,tang2021demon} as the attacker may not return the manipulated training data to the user or the attacker does not even interfere with the training data~\cite{liu2018trojaning,yao2019latent}.

\section{Conclusion}\label{sec:conclusion}
With the key insight that backdoor attacks have the intrinsic property of non-transferability, we present NTD to detect trigger inputs fed into a potentially backdoored MUT during the online phase as an easy-to-deploy plug-in defense mechanism. NTD can effectively detect trigger inputs targeting any class of an MUT using the global threshold. In addition, the per-class threshold can be trivially enabled to better protect security-sensitive classes that the attacker is more likely to target. To optimize its online detection latency, NTD retrieves all the features from the validation dataset in the offline phase for directly use during the online detection phase. For real-world scenarios allowing multiple trials, the NTD can further improve its online FRRs and FARs by enabling the multiple trials setting. 
As the reserved FE used by NTD is not affected by the backdoored (if exists) inserted into the MUT---this is especially true when the FE is provided by a reputable provider for highly repeatable and demanding tasks such as facial recognition in practice, NTD is agnostic to backdoor types and trigger types by design. Furthermore, NTD is user-friendly requiring neither ML expertise nor ML training, making itself distinguishable from existing backdoor countermeasures. 

\bibliographystyle{IEEEtran}
\bibliography{Reference}

\end{document}